\documentclass[11pt]{llncs}
\usepackage{graphicx}
\usepackage{cite}
\usepackage{url}
\usepackage{amssymb}
\def\R{\ensuremath{\mathbb R}}
\def\H{\ensuremath{\mathbb H}}
\def\Z{\ensuremath{\mathbb Z}}

\makeatletter
\def\hb@xt@{\hbox to }
\makeatother

\let\oldendproof\endproof
\def\endproof{\qed\oldendproof}

\begin{document}

\title{Quasiconvex Programming}
\author{David Eppstein}

\institute{Computer Science Department\\
Donald Bren School of Information \& Computer Sciences\\
University of California, Irvine\\
Irvine, CA 92697-3424\\
\email{eppstein@uci.edu}}

\date{ }

\maketitle

\begin{abstract}
We define {\em quasiconvex programming}, a form of generalized linear programming in which one seeks the point minimizing the pointwise maximum of a collection of quasiconvex functions.
We survey algorithms for solving quasiconvex programs either numerically or via generalizations of the dual simplex method from linear programming, and describe varied applications of this geometric optimization technique in meshing, scientific computation, information visualization, automated algorithm analysis, and robust statistics.
\end{abstract}

\pagestyle{plain}
\thispagestyle{empty}

\section{Introduction}

Quasiconvex programming is a form of geometric optimization, introduced by Amenta et al. in the context of mesh improvement techniques~\cite{AmeBerEpp-Algs-99} and since applied to other problems  in meshing, scientific computation, information visualization, automated algorithm analysis, and robust statistics~\cite{BerEpp-WADS-01-omt,BerEpp-SCG-03,Cha-SODA-04,Epp-SODA-04-qaba}.  If a problem can be formulated as a quasiconvex program of bounded dimension, it can be solved algorithmically in a linear number of constant-complexity primitive operations by generalized linear programming techniques, or numerically by generalized gradient descent techniques.  In this paper we survey quasiconvex programming algorithms and applications.

\subsection{Quasiconvex Functions}

Let $Y$ be a totally ordered set, for instance the real numbers $\R$ or integers $\Z$ ordered numerically.
For any function $f:X\mapsto Y$, and any value $\lambda\in Y$,
we define the {\em lower level set}
$$f^{\le\lambda}=\left\{x\in X\mid f(x)\le\lambda\right\}.$$
A function $q:X\mapsto Y$, where $X$ is a convex subset of $\R^d$, is called {\em quasiconvex}~\cite{DhaJoa-88} when its lower level sets are all convex.
A one-dimensional quasiconvex function is more commonly called {\em unimodal},
and another way to define a quasiconvex function is that it is unimodal along any line through its domain.

\begin{figure}[t]
\centering
\includegraphics[width=2.75in]{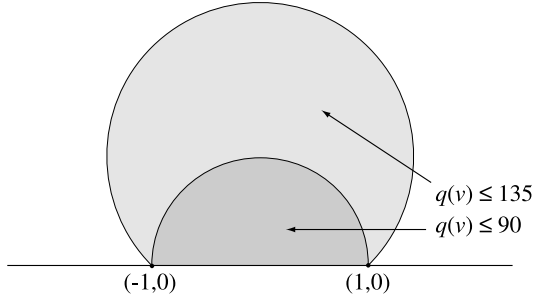}
\caption{Level sets of the quasiconvex function
$q(v)=180^\circ-\angle uvw$,
for $u=(-1,0)$ and $w=(1,0)$,
restricted to the halfplane $y\ge 0$.}
\label{fig:angles}
\end{figure}

As an example, let $H=\{(x,y)\mid y>0\}$ be the upper halfplane in $\R^2$,
let $u=(-1,0)$ and $w=(1,0)$, and let $q$ measure the angle complementary to the one
subtended by segment $uw$ from point $v$: $q(v)=180^\circ-\angle uvw$.  Then, each level set $q^{\le\lambda}$ consists of the intersection with $H$ of a disk having $u$ and $w$ on its boundary (Figure~\ref{fig:angles}).  Since these sets are all convex, $q$ is quasiconvex.

Quasiconvex functions are a generalization of the well-known set of {\em convex functions}, which are
the functions $\R^d\mapsto\R$ satisfying the inequality $f(p\bar x+(1-p)\bar y)\le p\,f(\bar x)+(1-p)f(\bar y)$
for all $\bar x,\bar y\in\R^d$ and all $0\le p\le 1$: it is a simple consequence of this inequality that any convex function has convex lower level sets.
However, there are many functions that are quasiconvex but not convex;
for instance, the complementary angle function $q$ defined above is not convex, as can be seen from the fact that its values are upper bounded by $180$.
As another example, the function $\chi_K(\bar x)$ that takes the value $0$ within a convex set $K$ and $1$ outside $K$
has as its lower level sets $K$ and $\R^d$, so is quasiconvex, but not convex.

If $r$ is convex or quasiconvex and $f:Y\mapsto Z$ is monotonically nondecreasing,
then $q(\bar x)=f(r(\bar x))$ is quasiconvex;
for instance the function $\chi_K$ above can be factored in this way into the composition of a convex function
$d_K(\bar x)$ measuring the Euclidean distance from $\bar x$ to $K$ with a monotonic function
$f$ mapping $0$ to itself and all larger values to~$1$.
In the other direction, given a quasiconvex function $q:X\mapsto Y$, one can often find a monotonic function $f:Y\mapsto\R$ that,
when composed with $q$, turns it into a convex function.
However this sort of convex composition is not always possible. For instance, in the case of the step function $\chi_K$ described above, any nonconstant composition of $\chi_K$ remains two-valued and hence cannot be convex.

\subsection{Nested Convex Families}

Quasiconvex functions are closely related to {\em nested convex families}.
Following Amenta et al.~\cite{AmeBerEpp-Algs-99}, we define a nested convex family to be a map
$\kappa:Y\mapsto K(\R^d)$, where $Y$ is a totally ordered set
and $K(\R^d)$ denotes the family of compact convex subsets of $\R^d$,
and where $\kappa$ is further required to satisfy the following two axiomatic requirements
(the second of which is a slight generalization of the original definition of Amenta et al.,
that allows $Y$ to be discrete):
\begin{enumerate}
\item For every $\lambda_1,\lambda_2\in Y$ with $\lambda_1<\lambda_2$,
$\kappa(\lambda_1)\subseteq\kappa(\lambda_2)$.
\item For all $\lambda\in Y$ for which $\lambda=\inf\{\lambda'\mid\lambda'>\lambda\}$, $\kappa(\lambda)=\bigcap_{\lambda'>\lambda}\kappa(\lambda')$.
\end{enumerate}

If $Y$ has the property that every subset of $Y$ has an infimum (for instance, $Y=\R\cup\{\infty,-\infty\}$), then
from any nested convex family $\kappa:Y\mapsto K(\R^d)$ we can define a function
$q_\kappa:\R^d\mapsto Y$ by the following formula:
$$q_\kappa(\bar x) = \inf\,\left\{\,\lambda \mid \bar x \in \kappa(\lambda)\,\right\}.$$

\begin{lemma}
\label{lem:qk-levels=k}
For any nested convex family $\kappa:Y\mapsto K(\R^d)$ and any $\lambda\in Y$,
$q_\kappa^{\le\lambda}=\kappa(\lambda)$.
\end{lemma}

\begin{proof}
The lower level sets of $q_\kappa$ are
$$q_\kappa^{\le\lambda}
=\bigl\{\bar x\in R^d\mid q_\kappa(\bar x)\le\lambda\bigl\}
=\bigl\{\bar x\in R^d\mid 
\inf\,\left\{\,\lambda' \mid \bar x \in \kappa(\lambda')\,\right\}
\le\lambda\bigl\}.
$$
For any $\bar x\in\kappa(\lambda)$,
$\lambda\in\left\{\,\lambda' \mid \bar x \in \kappa(\lambda')\,\right\}$
so the infimum of this set can not be greater than $\lambda$ and $\bar x\in q_\kappa^{\le\lambda}$.
For any $\bar x\notin\kappa(\lambda)$,
$\inf\,\left\{\,\lambda' \mid \bar x \in \kappa(\lambda')\,\right\}\ge\lambda^+>\lambda$
by the second property of nested convex families,
so $\bar x\notin q_\kappa^{\le\lambda}$.
Therefore, $q_\kappa^{\le\lambda}=\kappa(\lambda)$.
\end{proof}

In particular, $q_\kappa$ has convex lower level sets and so is quasiconvex.

Conversely, suppose that $q$ is quasiconvex and has bounded lower level sets.
Then we can define a nested convex family
$$\kappa_q(\lambda)=\left\{
\begin{array}{ll}
\bigcap_{\lambda'>\lambda}\mathop{\rm cl}(q^{\le\lambda'})&\hbox{\quad if $\lambda=\inf\{\lambda'\mid\lambda'>\lambda\}$}\\[0.75ex]
\mathop{\rm cl}(q^{\le\lambda})&\hbox{\quad otherwise}
\end{array}
\right.$$
where cl denotes the topological closure operation.

If $q$ does not have bounded lower level sets, we can still form a nested convex family
by restricting our attention to a compact convex subdomain $K\subset\R^d$:
$$\kappa_{q,K}(\lambda)=\left\{
\begin{array}{ll}
\bigcap_{\lambda'>\lambda}\mathop{\rm cl}(K\cap q^{\le\lambda'})&\hbox{\quad if $\lambda=\inf\{\lambda'\mid\lambda'>\lambda\}$}\\[0.75ex]
\mathop{\rm cl}(K\cap q^{\le\lambda})&\hbox{\quad otherwise}
\end{array}
\right.$$
This restriction to a compact subdomain is necessary to handle linear functions and other functions without bounded level sets within our mathematical framework.

The following two theorems allow us to use nested convex families and quasiconvex functions
interchangeably for each other for most purposes: more specifically, a nested convex family conveys exactly the same information as a continuous quasiconvex function with bounded lower level sets.  Thus, later, we will use whichever of the two notions is more convenient for the purposes at hand, using these theorems to replace an object of one type for an object of the other in any algorithms or lemmas needed for our results.

\begin{theorem}
For any nested convex family $\kappa$, $\kappa=\kappa_{q_\kappa}$.
\end{theorem}

\begin{proof}
If $\lambda$ is not an infimum of larger values,
then $q_\kappa(x)\le\lambda$ if and only if 
$x\in \kappa(\lambda)$.
So
$\kappa_{q_\kappa}(\lambda)=
\mathop{\rm cl}({q_\kappa}^{\le\lambda})=
\{x\mid q_\kappa(x)\le\lambda\}=\kappa(\lambda)$.

Otherwise, by Lemma~\ref{lem:qk-levels=k},
$$\kappa_{q_\kappa}(\lambda)=\bigcap_{\lambda'>\lambda}\mathop{\rm cl}(\kappa(\lambda'))$$
The closure operation does not modify the set $\kappa(\lambda')$, because it is already closed,
so we can replace $\mathop{\rm cl}(\kappa(\lambda'))$ above by $\kappa(\lambda'))$,
giving
$$\kappa_{q_\kappa}(\lambda)=\bigcap_{\lambda'>\lambda}\kappa(\lambda').$$
The intersection on the right hand side of the equation
further simplifies to $\kappa(\lambda)$ by the second property of nested convex families.
\end{proof}

\begin{theorem}
If $q:X\mapsto\R$ is a continuous quasiconvex function with bounded lower level sets, then $q_{\kappa_q}=q$.
\end{theorem}

\begin{proof}
By Lemma~\ref{lem:qk-levels=k},
$q_{\kappa_q}^{\le\lambda}=\kappa_q(\lambda)$.
Assume first that $\lambda=\inf\{\lambda'\mid\lambda'>\lambda\}$.
Expanding the definition of $\kappa_q$, we get
$$
q_{\kappa_q}^{\le\lambda}=
\bigcap_{\lambda'>\lambda}\mathop{\rm cl}(q^{\le\lambda'}).
$$
If $q$ is continuous, its level sets are closed, so we can simplify this to
$$
q_{\kappa_q}^{\le\lambda}=
\bigcap_{\lambda'>\lambda}q^{\le\lambda'}.
$$
Suppose the intersection on the right hand side of the formula is nonempty, and let $\bar x$ be any point in this intersection.  We wish to show that $q(\bar x)\le\lambda$,
so suppose for a contradiction that $q(\bar x)>\lambda$.
But then there is a value $\lambda'$ strictly between $\lambda$ and $q(\bar x)$
(else $\lambda$ would not be the infimum of all greater values),
and $\bar x\notin q^{\le\lambda'}$, contradicting the assumption that $\bar x$ is in the intersection.  Therefore, $q(\bar x)$ must be at most equal to $\lambda$.

As we have now shown that
$q(\bar x)\le\lambda$ for any $\bar x$ in $q_{\kappa_q}^{\le\lambda}$, it follows that $q_{\kappa_q}^{\le\lambda}$ can
not contain any points outside $q^{\le\lambda}$.
On the other hand, $q_{\kappa_q}^{\le\lambda}$
is formed by intersecting a collection of supersets of $q^{\le\lambda}$,
so it contains all points inside $q^{\le\lambda}$.
Therefore, the two sets are equal.

If $\lambda\ne\inf\{\lambda'>\lambda\}$,
the same equality can be seen even more simply to be true, since we have no intersection operation to eliminate.
Since $q_{\kappa_q}$ and $q$ have the same level sets, they are the same function.
\end{proof}

Due to these two theorems, we do not lose any information by using the function $q_\kappa$ in place of the nested convex family $\kappa$, or by using the nested convex family $\kappa_{q_\kappa}=\kappa$ in place of a quasiconvex function that is of the form $q=q_\kappa$ or in place of a continous quasiconvex function with bounded lower level sets. In most situations quasiconvex functions and nested convex families can be treated as equivalent and interchangeable: if we are given a quasiconvex function $q$ and need a nested convex family, we can use the family $\kappa_q$, and if we are given a nested convex family $\kappa$ and need a quasiconvex function, we can use the function $q_\kappa$ or $q_{\kappa,K}$.
Our quasiconvex programs'  formal definition will involve inputs that are nested convex families only,
but in our applications of quasiconvex programming we will describe inputs that are quasiconvex functions, and which will be assumed to be converted to nested convex families as described above.

\subsection{Quasiconvex Programs}
\label{sec:qcp}

If a finite set of functions $q_i$ are all quasiconvex and have the same domain and range, then the function $Q(\bar x)=\max_{i\in S} q_i(\bar x)$ is also quasiconvex, and it becomes of interest to find a point where $Q$ achieves its minimum value.  For instance, in Section~\ref{sec:smallball} below we discuss in more detail the smallest enclosing ball problem, which can be defined by a finite set of functions $q_i$, each of which measures the distance to an input site; the minimum of $Q$ marks the center of the smallest enclosing ball of the sites.
Informally, we use {\em quasiconvex programming} to describe this search for the point minimizing the pointwise maximum of a finite set of quasiconvex functions\footnote{The term {\em quasiconvex programming} has also been applied to the problem of minimizing a single quasiconvex function over a convex domain; e.g., see \cite{Kiw-MP-01,Xu-JOTA-01}. The two formulations are easily converted to each other using the ideas described in Section~\ref{sec:linconq}.  For the applications described in this survey, we prefer the formulation involving minimizing the pointwise maximum of multiple quasiconvex functions, as it places greater emphasis on combinatorial algorithms and less on numerical optimization}.

More formally, Amenta et al.~\cite{AmeBerEpp-Algs-99} originally defined a {\em quasiconvex program} to be formed by a set of nested convex families
$S=\{\kappa_1,\kappa_2,\ldots \kappa_n\}$; the task to be solved is
finding the value
$$\Lambda(S)=
\inf\Big\{\,
                (\lambda,\bar x) \Bigm{|}
                                \bar x\in \mathop{\textstyle\bigcap}\limits_{\kappa_i\in S}\kappa_i(\lambda)
\Big\}
$$
where the infimum is taken in the lexicographic ordering,
first by $\lambda$ and then by the coordinates of~$\bar x$.
However, we can simplify the infimum operation in this definition by replacing it with a minimum;
that is, it is always true that the set defined on the right hand side of the definition has a least point $\Lambda(S)$.
To prove this, suppose that $(\lambda,\bar x)$ is the infimum, that is, there is a sequence of pairs $(\lambda_j,\bar x_j)$ in the right hand side intersection that converges
to $(\lambda,\bar x)$, and $(\lambda,\bar x)$ is the smallest pair with this property.
Clearly, each $\lambda_j\ge\lambda$ (else $(\lambda_j,\bar x_j)$ would be a better solution)
and it follows from the fact that the sets $\kappa_i$ are closed and nested that
we can take each $\bar x_j=\bar x$.
But then, it follows from the second property of nested convex families that $\bar x\in\kappa_i(\lambda)$ for all $\kappa_i\in S$.

In terms of the quasiconvex functions defining a quasiconvex program, we would like to say that the value of the program consists of a pair $(\lambda,\bar x)$ such that, for each input function $q_i$, $q_i(\bar x)\le\lambda$, and that no other pair with the same property has a smaller value of~$\lambda$.  However, $\max_i q_i(\bar x)$ may not equal $\lambda$ if at least one of the input quasiconvex functions is discontinuous.
For instance, consider a one-dimensional quasiconvex program with two functions $q_0(x)=|x|$,
$q_1(x)=1$ for $x\ge0$, and $q_1(x)=0$ for $x<0$. This program has value $(0,0)$, but 
$\max\{q_0(0),q_1(0)\}=1$.
The most we can say in general is that there exists a sequence of points $\bar x_j$ converging to $x$ with
$\lim_{j\rightarrow\infty}\max_i q_i(\bar x_j)=\lambda$.  This technicality is, however, not generally a problem in our applications.

In subsequent sections we explore various examples of quasiconvex programs,
algorithms for quasiconvex programming, and applications of those algorithms.

\section{Examples}

We begin our study of quasiconvex programming by going through some simple examples of geometric optimization problems, and showing how they may be formulated as low-dimensional quasiconvex programs.

\subsection{Sighting point}
\label{sec:sighting}

When we introduced the definition of quasiconvex functions, we used as an example
the complementary angle subtended by a line segment from a point: $q(v)=180^\circ-\angle uvw$.
If we have a collection of line segments forming a star-shaped polygon,
and form a quasiconvex program from the functions corresponding to each line segment,
then the point $v$ that minimizes the maximum function value must lie in the kernel of the polygon.
If we define the {\em angular resolution} of the polygon from $v$ to be the minimum angle formed by any two consecutive vertices as seen from $v$, then this choice of $v$ makes the angular resolution be as large as possible.

This problem of maximizing the angular resolution was used by Matou{\v{s}}ek et al.~\cite{MatShaWel-Algo-96} as an example of an LP-type problem that does not form a convex program.
It can also be viewed as a special case of the mesh smoothing application described below in Section~\ref{sec:meshsmooth}.

Earlier, McKay~\cite{McK-89} had asked about a similar problem in which one wishes to choose a viewpoint maximizing the angular resolution of an unordered set of points that is not connected into a star-shaped polygon.
However, it does not seem possible to form a quasiconvex program from this version of the problem: for star-shaped polygons, we know on which side of each line segment the optimal point must lie, so we can use quasiconvex functions with level sets that are intersections of disks and halfplanes, but for point sets, without knowing where the viewpoint lies with respect to the line through any pair of points, we need to use the absolute value $|q(v)|$ of the angle formed at $v$ by each pair of points. This modification leads to non-quasiconvex functions with level sets that are unions or intersections of two disks.  It remains open whether an efficient algorithm for McKay's sighting point problem exists.

\subsection{Smallest Enclosing Ball}
\label{sec:smallball}

\begin{figure}[t]
\centering
\includegraphics[scale=0.4]{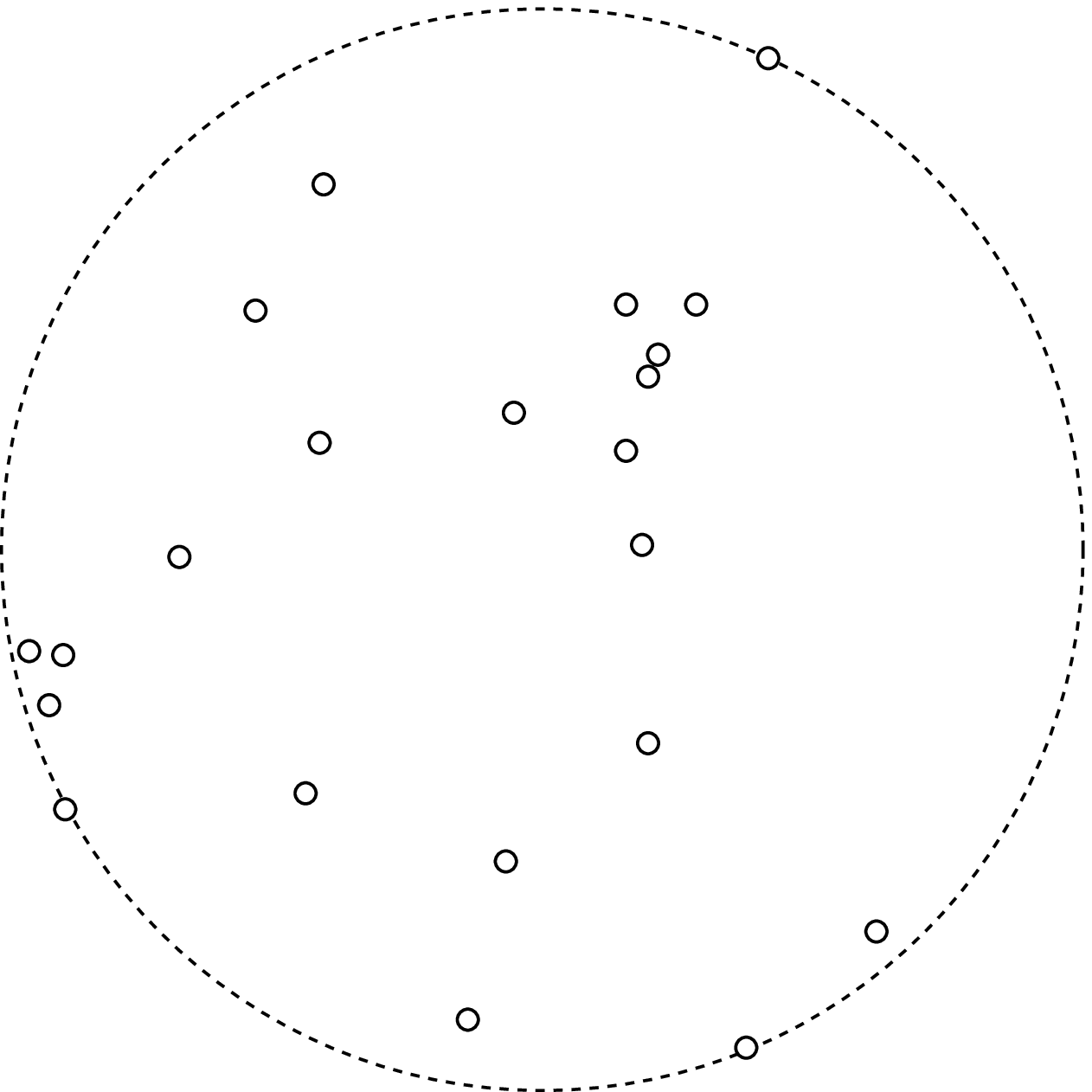}
\qquad
\includegraphics[scale=0.4]{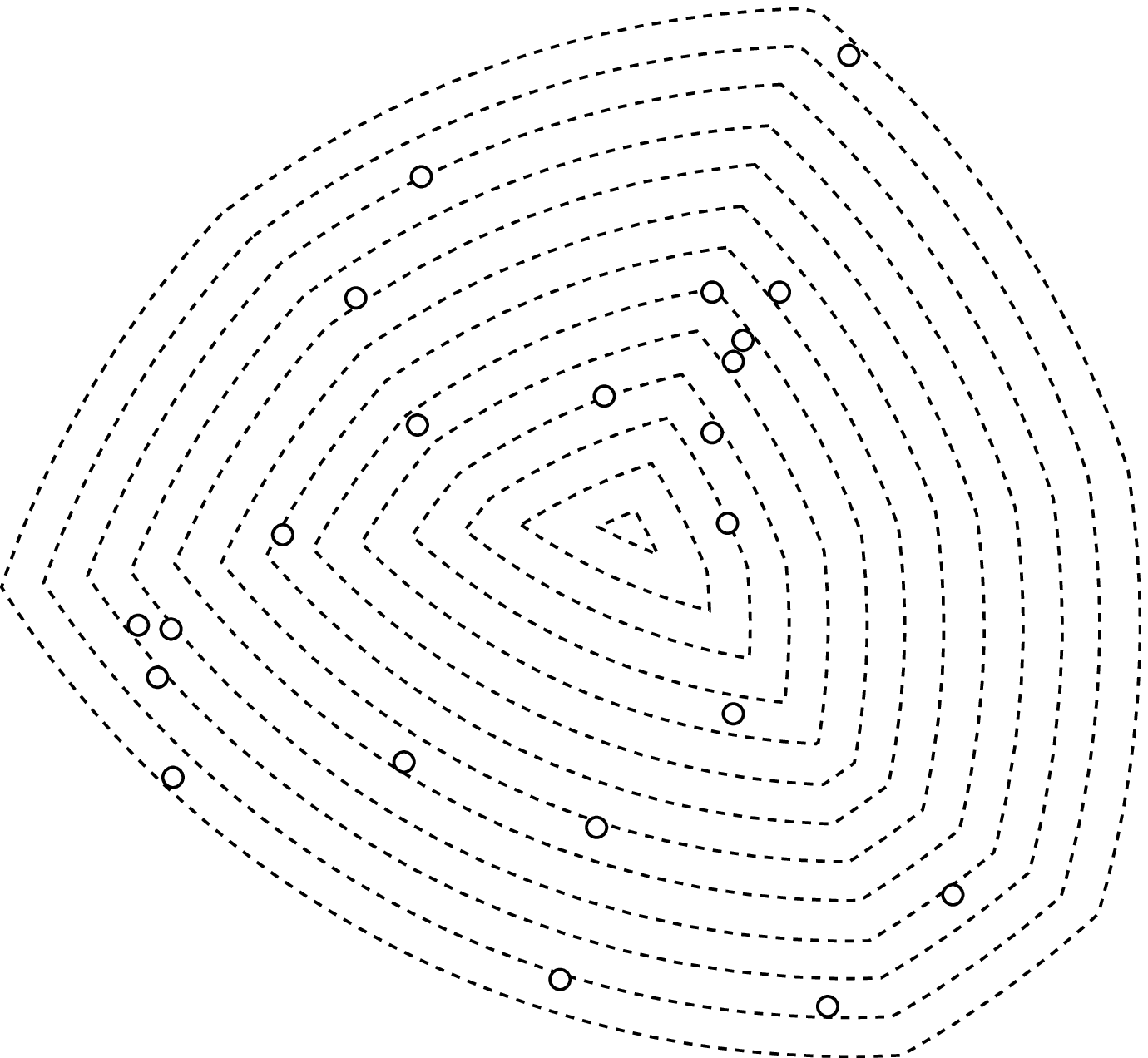}
\caption{Smallest enclosing ball of a set of points (left), and the level sets of $\max_i q_i(x)$ for the distance functions $q_i$ defining the quasiconvex program for the smallest enclosing ball (right).}
\label{fig:seb}
\end{figure}

Consider the problem of finding the minimum radius Euclidean sphere that encloses all of a set of points $S=\{\bar p_i\}\subset\R^d$ (Figure~\ref{fig:seb}, left).  As we show below, this smallest enclosing ball problem can easily be formulated as a quasiconvex program.  The smallest enclosing ball problem has been well studied and linear time algorithms are known in any fixed dimension~\cite{Dye-SJC-84,FisGaeKut-ESA-03,Gae-ESA-99,Meg-SJC-83,Wel-NRNTCS-91}, so the quasiconvex programming formulation does not lead to improved solutions for this problem, but it provides an illuminating example of how to find such a formulation more generally, and in later sections we will use the smallest enclosing ball example to illustrate our quasiconvex programming algorithms.

Define the function $q_i(\bar x)=d(\bar x,\bar p_i)$ where $d$ is the Euclidean distance.  Then the level set $q_i^{\le\lambda}$
is simply a Euclidean ball of radius $\lambda$ centered at $\bar p_i$, so $q_i$ is quasiconvex (in fact, it is convex).  The function $q_S(\bar x)=\max_i q_i(\bar x)$ (the level sets of which are depicted in Figure~\ref{fig:seb}, right) measures the maximum distance from $\bar x$ to any of the input points, so a Euclidean ball of radius $q_S(\bar x)$ centered at $\bar x$ will enclose all the points and is the smallest ball centered at $\bar x$ that encloses all the points.

If we form a quasiconvex program from the functions $q_i$, the solution to the program consists
of a pair $(\lambda,\bar x)$ where $\lambda=q_S(\bar x)$ and $\lambda$ is as small as possible.  That is, the ball with radius $\lambda$ centered at $\bar x$ is the smallest enclosing ball of the input points.

Any smallest enclosing ball problem has a {\em basis} of at most $d+1$ points that determine its value.
More generally, it will turn out that any quasiconvex program's value is similarly determined by
a small number of the input functions; this phenomenon will prove central in our ability to apply generalized linear programming algorithms to solve quasiconvex programs.

If we generalize each $q_i$ to be the Euclidean distance to a convex set $K_i$,
the resulting quasiconvex program finds the smallest sphere that touches or encloses each $K_i$.
In a slightly different generalization, if we let $q_i(\bar x)=d(\bar x,\bar p_i)+r_i$,
a sphere centered at $\bar x$ with radius $q_i(\bar x)$ or larger will contain the sphere centered
at $\bar p_i$ with radius $r_i$.  So, solving the quasiconvex program with this family of functions $q_i$ will find the smallest enclosing ball of a family of balls~\cite{Meg-DCG-89,GaeFis-SCG-03}.

\subsection{Hyperbolic Smallest Enclosing Ball}

Although we have defined quasiconvex programming in terms of Euclidean space $\R^n$,
the definition involves only concepts such as convexity that apply equally well to other geometries such as hyperbolic space $\H^n$.  Hyperbolic geometry (e.g. see~\cite{Ive-92}) may be defined in various ways; for instance by letting $\H^n$ consist of the unit vectors of $\R^{n+1}$ according to the
inner product $\left<\bar x,\bar y\right>=\sum_{i<n}(x_iy_i)-x_ny_n$,
and defining the distance $d(\bar x,\bar y)=\cosh^{-1}\left<\bar x,\bar y\right>$.
Angles, congruence, lines, hyperplanes, and other familiar Euclidean concepts can also be defined in a straightforward way for hyperbolic space.
Hyperbolic geometry satisfies many of the same axioms as Euclidean geometry, but not the famous {\em parallel postulate}: in the hyperbolic plane $\H^2$, given a line $\ell$ and a point $p\notin\ell$, there will be infinitely many lines through $p$ that do not meet $\ell$.
A hyperbolic convex set $K$ is defined as in Euclidean space to be one in which, for any two points $\{p,q\}\subset K$, all points on the line segment connecting $p$ to $q$ also belong to $K$.
Similarly, a quasiconvex function $\H^n\mapsto \R$ is one for which all lower level sets are convex,
or equivalently one that is unimodal on any line in $\H^n$.
As in the Euclidean case we may define a hyperbolic quasiconvex program to be the problem of searching for the point minimizing the pointwise maximum of a collection of hyperbolic quasiconvex functions.

\begin{figure}[t]
\centering
\includegraphics[width=2.25in]{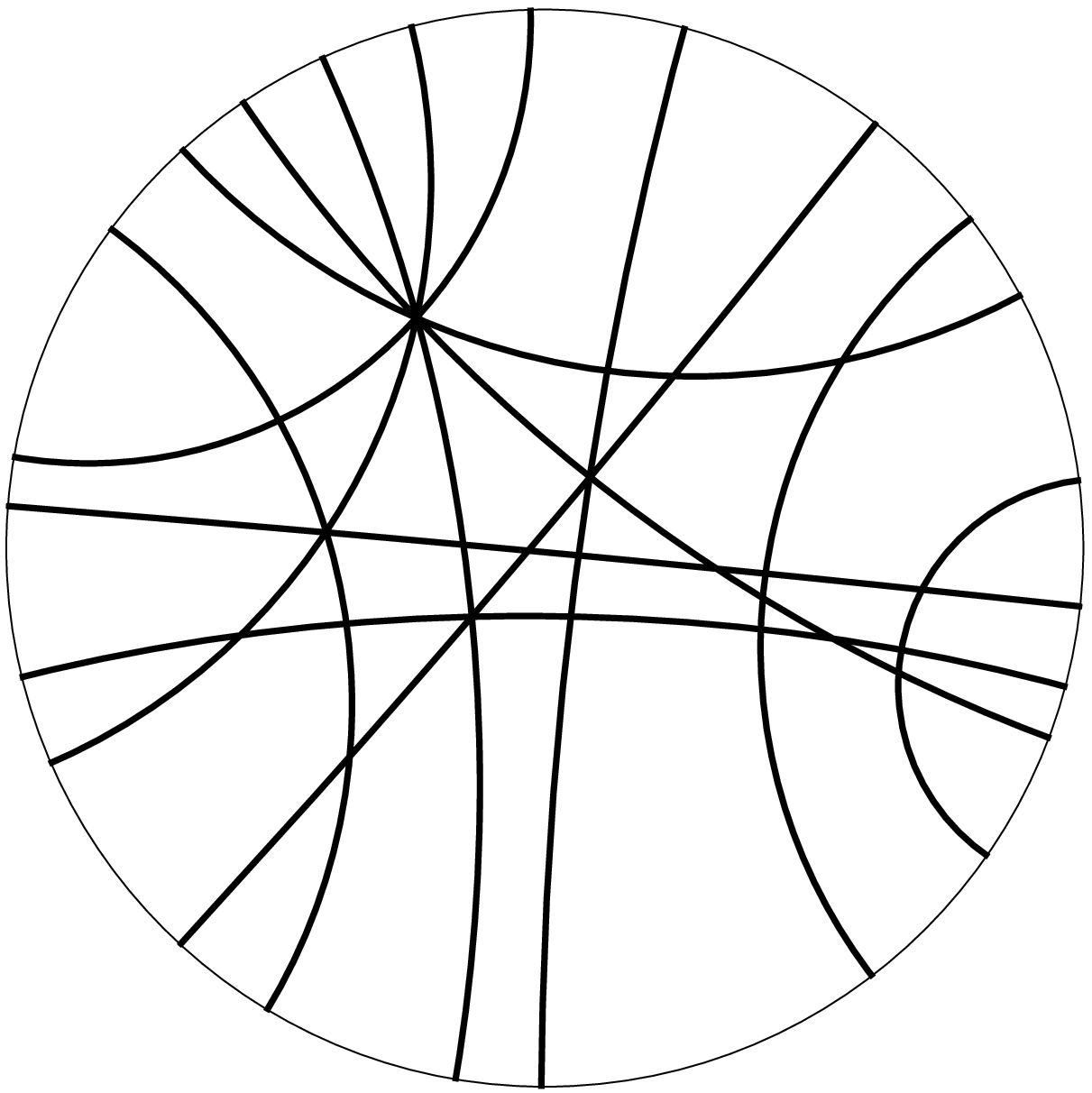}
\qquad
\includegraphics[width=2.25in]{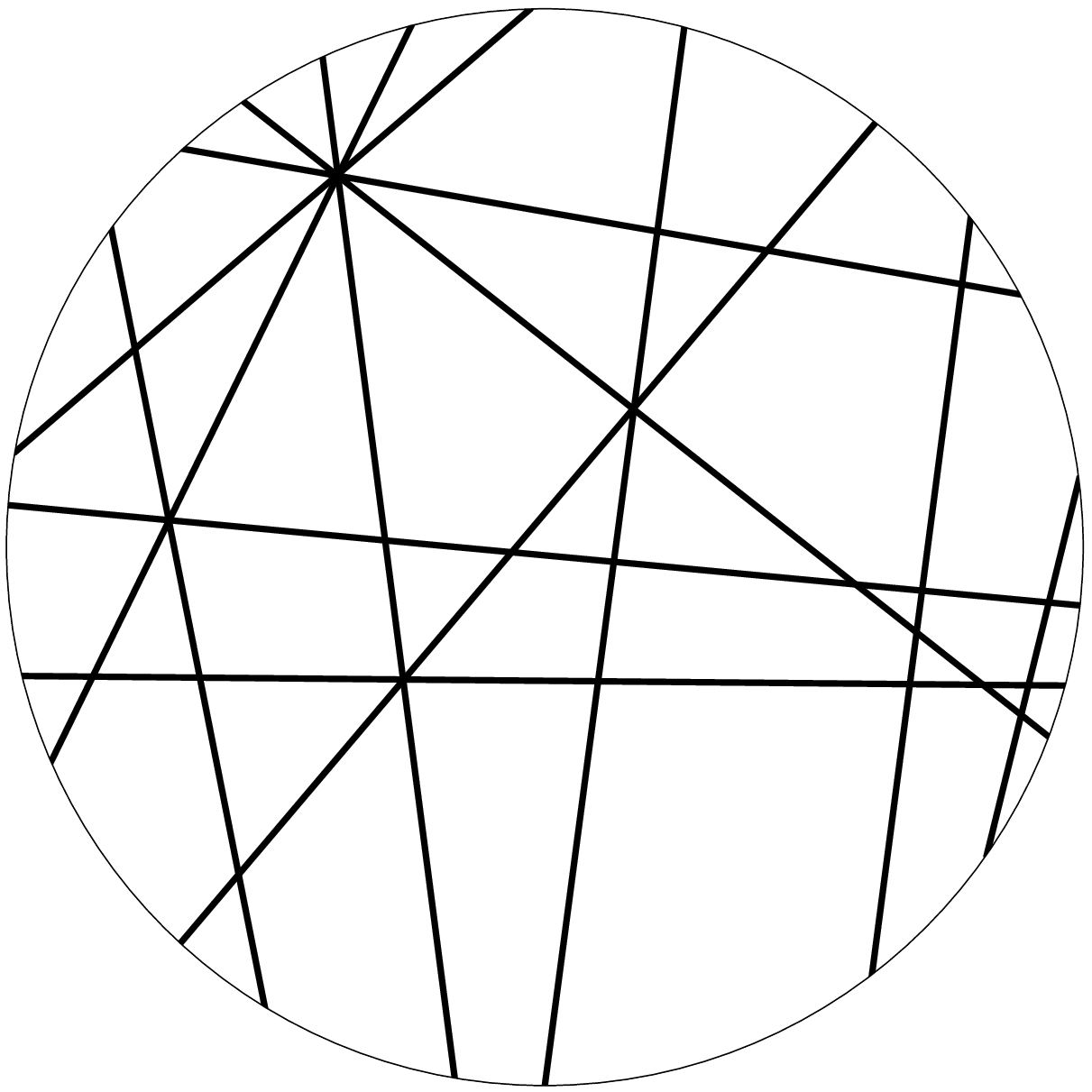}
\caption{Poincar\'e (left) and Klein (right) models of the hyperbolic
plane. Both models show the same hyperbolic arrangement of lines; analogous models exist for any higher dimensional hyperbolic space.  Figure taken from~\cite{BerEpp-WADS-01-omt}.}
\label{fig:poinklein}
\end{figure}

There are several standard ways of representing the points and other geometric objects of Hyperbolic space within a Euclidean space, of which the two best known are the Poincar\'e and Klein models
(Figure~\ref{fig:poinklein}).  In the Poincar\'e model, the points of $\H^n$ are represented as Euclidean points interior to an $n$-dimensional unit ball or halfspace, and lines of $\H^n$ are represented as arcs of circles that meet the boundary of this unit ball or halfspace perpendicularly.  In this model, the hyperbolic angle between two objects in $\H^n$ is equal to the Euclidean angle between the models of those objects, and hyperbolic circles and spheres are modeled by Euclidean circles and spheres; however, hyperbolic distances do not equal distances within the Poincar\'e model, and objects that are straight or flat hyperbolically may have curved models.  In the Klein model, again, points of $\H^n$ are represented as Euclidean points interior to an $n$-dimensional unit ball, but the hyperbolic line connecting two points is represented as the restriction to the ball of the Euclidean line connecting the models of those points.  In this model, angles and distances may be distorted but straightness is preserved: a straight or flat hyperbolic object will have a straight or flat model.
In particular, since the definition of convexity involves only straight line segments,
a convex hyperbolic object will have a convex Klein model and vice versa.
The Poincar\'e and Klein models for a hyperbolic space are not uniquely defined, as one may choose
any hyperbolic point to be modeled by the center of the Euclidean unit ball, and that ball may rotate arbitrarily around its center.

If we let $k$ be a function mapping $\H^n$ to a Klein model in $\R^n$,
and if each $q_i(\bar x)$ is a hyperbolic quasiconvex function, then
$\hat q_i(\bar x)=q_i(k^{-1}(\bar x))$ is a Euclidean quasiconvex function.
More, $\hat q_i$ has bounded lower levels sets since they are all subsets of the unit ball.
Let $(\lambda,\bar x)$ be the solution to the Euclidean quasiconvex program defined by the set of functions $\hat q_i$.  Then, if $\bar x$ is interior to the unit ball defining the Klein model,
$(\lambda,k^{-1}(\bar x))$ is the solution to the hyperbolic quasiconvex program
defined by the original functions $q_i$.  On the other hand, $\bar x$ may be
on the boundary of the Klein model; if so, $\bar x$ may be viewed as an infinite point of the hyperbolic space, and is the limit of sequence of points within the space with monotonically decreasing values.
The latter possibility, of an infinite solution to the quasiconvex program, can only occur if some of the hyperbolic quasiconvex functions have unbounded lower level sets.
Therefore, as Bern and Eppstein~\cite{BerEpp-WADS-01-omt} noted,
hyperbolic quasiconvex programs may in general be solved as easily as their Euclidean counterparts.

As an example, consider the problem of finding the hyperbolic ball of minimum radius containing all of a collection of hyperbolic points $\bar p_i$.  As in the Euclidean case, we can define $q_i(\bar x)$ to be the (hyperbolic) distance from $\bar x$ to $\bar p_i$; this function has convex hyperbolic balls as its level sets, so it is quasiconvex.  And, just as in the Euclidean case, the solution to the quasiconvex program defined by the functions $q_i$ is the pair $(\lambda,\bar x)$ where the hyperbolic ball of radius $\lambda$  centered at $\bar x$ is the minimum enclosing ball of the points $\bar p_i$.

\subsection{Optimal Illumination}

Suppose that we have a room (modeled as a possibly nonconvex three-dimensional polyhedron) and wish to place a point source of light in order to light up the whole room as brightly as possible: that is, we wish to maximize the minimum illumination received on any point of the room's surface.  The quasiconvex programs we are studying solve min-max rather than max-min problems, but that is easily handled by negating the input functions.

So, let $q_i(\bar x)$ be the negation of the intensity of light received at point $i$ of the room's surface, as a function of $\bar x$, the position of the light source.  It is not hard to see that, within any face of the polyhedron, the light intensity is least at some vertex of the face, since those are the points at maximal distance from the light source and with minimal angle to it.  Therefore, we need only consider a finite number of possibilities for $i$: one for each pair $(f,v)$ where $f$ is a face of the polyhedron and $v$ is a vertex of $f$.  For each such pair, we can compute $q_i$ via a simple formula of optics, $q_i(\bar x)=-\bar u\cdot(\bar x-v)/d(\bar x,v)^3$, where $d$ is as usual the Euclidean distance, and $u$ is a unit vector facing inwards at a perpendicular angle to $f$.
In this formula, one factor $\bar u\cdot(\bar x-v)/d(\bar x,v)$ accounts for the angle of incidence of light from the source onto the part of face $v$ near vertex $f$, while the other factor $1/d(\bar x,v)^2$ accounts for the inverse-square rule for falloff of light from a point source in three-dimensional space.
Note that we can neglect occlusions from other faces in this formula, because, if some face is occluded, then at least one other face will be facing away from the light source and entirely unilluminated; this unilluminated face will dominate the occluded one in our min-max optimization.

In \cite{AmeBerEpp-Algs-99}, as part of a proof of quasiconvexity of a more complex function used for smoothing three-dimensional meshes by solid angles, we showed that the function $q_i$ defined above is quasiconvex; more precisely, we showed that $(-q_i(\bar x))^{-1/2}$ is a convex function of $\bar x$ by using {\em Mathematica} to calculate the principal determinants of its Hessian, and by showing from the structure of the resulting formulae that these determinants are always nonnegative.  Therefore, we can express the problem of finding an optimal illumination point as a quasiconvex program.

\subsection{Longest Intersecting Prefix}
\label{sec:lip}

\begin{figure}[t]
\centering
\includegraphics[width=3in]{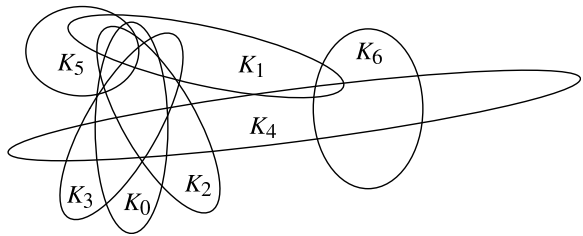}
\caption{Instance of a longest intersecting prefix problem. The longest intersecting prefix is $(K_0,K_1,K_2,K_3)$.}
\label{fig:IntersectingPrefix}
\end{figure}

This example is due to Chan~\cite{Cha-SODA-04}.
Suppose we are given an ordered sequence of convex sets $K_i$, $0\le i<n$, that are all subsets of the same compact convex set $X\subset\R^d$.  We would like to find the maximum value $\ell$ such that
$\cap_{i<\ell}K_i\neq\emptyset$.  That is, we would like to find the longest prefix of the input sequence, such that the convex sets in this prefix have a nonempty intersection
(Figure~\ref{fig:IntersectingPrefix}).

To represent this as a quasiconvex program, define a nested convex family $\kappa_i:\Z\mapsto K(\R^d)$ for each set $K_i$ in the sequence, as follows:
$$
\kappa_i(\lambda)=\left\{
\begin{array}{ll}
K_i,&\hbox{\quad if $\lambda<-i$}\\
X,&\hbox{\quad otherwise.}
\end{array}
\right.
$$
The optimal value $(\lambda,\bar x)$ for the quasiconvex program formed by this set of nested convex families has
$\bar x\in\kappa_i(\lambda)=K_i$ for all $i<-\lambda$,
so the prefix of sets with index up to (but not including) $-\lambda$ has a nonempty intersection containing $\bar x$.  Since the quasiconvex program solution minimizes $\lambda$, $-\lambda$ is the maximum value with this property.
That is, the first $-\lambda$ values of the sequence $K_i$ form its longest intersecting prefix.

More generally, the same technique applies equally well when each of the convex sets $K_i$ has an associated value $k_i$, and we must find the maximum value $\ell$ such that
$\cap_{k_i<\ell}K_i\neq\emptyset$.  The longest intersecting prefix problem can be seen as a special case of this problem in which the values $k_i$ form a permutation of the integers from $0$ to $n-1$.  We will see an instance of this generalized longest intersecting prefix problem, in which the values $k_i$ are integers with some repeated values, when we describe Chan's solution to the Tukey median problem.

\subsection{Linear, Convex, Quasiconvex}
\label{sec:linconq}

\begin{figure}[t]
\centering
\includegraphics[width=3in]{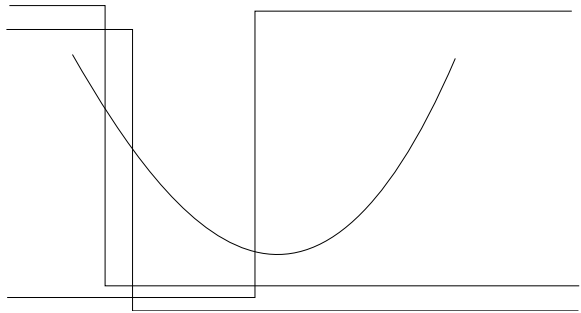}
\caption{Conversion of convex program into quasiconvex program, by treating each halfspace constraint as a quasiconvex step function.}
\label{fig:ConvexQuasiconvex}
\end{figure}

There are many ways of modeling linear programs, but one of the simplest is the following:
a linear program is the search for a vector $\bar x$ that satisfies all of a set of closed linear inequalities
$\bar a_i\cdot\bar x\ge b_i$ and that, among all such feasible vectors, minimizes a linear objective function $f(x)=\bar c\cdot\bar x$.  The vectors $\bar x$, $\bar a_i$, and $\bar c$ all have the same dimension, which we call the dimension of the linear program.
We typically use the symbol $n$ to denote the number of inequalities in the linear program.
It is often useful to generalize such programs somewhat, by keeping the linear constraints but allowing the objective function $f(x)$ to be convex instead of linear; such a generalization is known as a convex program, and many linear programming algorithms can be adapted to handle the convex case as well.

For instance, consider the following geometric problem, which arises in collision detection algorithms for maintaining simulations of virtual environments: we are given as input two $k$-dimensional convex bodies $P$ and $Q$, specified as intersections of halfspaces $P=\cap P_i$ and $Q=\cap Q_i$; we wish to find the closest pair of points $\bar p,\bar q$ with $\bar p\in P$ and $\bar q\in Q$.
If we view $\bar p,\bar q$ as forming a $2k$-dimensional vector $\bar x$, then each constraint
$\bar p\in P_i$ or $\bar q\in Q_i$ is linear in $\bar x$, but the objective function $d(\bar p,\bar q)$ is nonlinear: evaluating the distance using the Pythagorean formula results in a formula that is the square root of a sum of squares of differences of coordinates.  We can square the formula to eliminate the square root, but what remains is a convex quadratic function.  Thus, the closest distance problem can be expressed as a convex program; similar formulations are also possible when $P$ and $Q$ are expressed as convex hulls of their vertex sets~\cite{MatShaWel-Algo-96}.

These formulations seem somewhat different from our quasiconvex programming framework: in the linear and convex programming formulations above, we have a large set of constraints and a single objective function, while in quasiconvex programming we have many input functions that take a role more analogous to objectives than constraints.  Nevertheless, as we now show, any linear or convex program can be modeled as  a quasiconvex program.  Intuitively, the idea is simply to treat each halfspace constraint as a quasiconvex step function, and include them together with the convex objective functions in the set of quasiconvex functions defining a quasiconvex program
(Figure~\ref{fig:ConvexQuasiconvex}).

\begin{theorem}
Suppose a convex program is specified by $n$ linear inequalities $\bar a_i\cdot\bar x\ge b_i$
and a convex objective function $f(\bar x)$, and suppose that the solution of this convex program is known to lie within a compact convex region $K$.  Then we can find a set of $n+1$ nested convex families $\kappa_i(\lambda)$ such that the solution $(\lambda,\bar x)$ of the quasiconvex program formed by these nested convex families is an optimal solution to the convex program, with $\lambda=f(\bar x)$.
\end{theorem}

\begin{proof}
For each inequality $\bar a_i\cdot\bar x\ge b_i$ form a nested convex family
$\kappa_i(\lambda)=K\cap\{\bar x\mid\bar a_i\cdot\bar x\ge b_i\}$; that is, $\kappa_i$ ignores its argument $\lambda$ and produces a constant compact convex set of the points satisfying the $i$th inequality.  Also form a nested convex family
$\kappa_n=\kappa_{f,K}$ representing the objective function.

If $(\lambda,\bar x)$ is the optimal solution to the quasiconvex program defined by the nested convex families $\kappa_i$, then $\bar a_i\cdot\bar x\ge b_i$ (else $\bar x$ would not be contained in
$\kappa_i(\lambda)$) and $\lambda=f(\bar x)$ (else either $\bar x$ would be outside $\kappa_n(\lambda)$ or the pair $(f(\bar x),\bar x)$ would be a better solution).
There could be no $\bar y$ satisfying all constraints $\bar a_i\cdot\bar y\ge b_i$
with $f(\bar y)<\lambda$, else $(f(\bar y),\bar y)$ would be a better solution than
$(\lambda,\bar x)$ for the quasiconvex program.  Therefore, $\bar x$ provides the optimal solution to the convex program as the result states.
\end{proof}

The region $K$ is needed for this result as a technicality, because our quasiconvex programming formulation requires the nested convex families to be compact.  In practice, though, it is not generally difficult to find $K$;
for instance, in the problem of finding closest distances between convex bodies, we could let $K$ be a bounding box defined by extreme points of the convex bodies in each axis-aligned direction.

\section{Algorithms}

We now discuss techniques for solving quasiconvex programs, both numerically and combinatorially.

\subsection{Generalized Linear Programming}

Although linear programs can be solved in polynomial time, regardless of dimension~\cite{Kar-Comb-84,Kha-CMMP-80}, known results in this direction involve time bounds that depend not just on the number and dimension of the constraints, but also on the magnitude of the coordinates used to specify the constraints.
In typical computational geometry applications the dimension is bounded but these magnitudes may not be, so there has been a long line of work on linear programming algorithms that take a linear amount of time in terms of the number of constraints, independent of the magnitude of coordinates,
but possibly with an exponential dependence on the dimension of the
problem~\cite{AdlSha-MP-93,ChaMat-SODA-93,Cla-IPL-86,Cla-DCG-87,Cla-JACM-95,DyeFri-MP-89,MatShaWel-Algo-96,Meg-SJC-83,Meg-JACM-84,Sei-DCG-91}.
In most cases, these algorithms can be interpreted as {\em dual simplex methods}: as they progress, they maintain a {\em basis} of $d$ constraints, and the point $\bar x$ optimizing the objective function subject to the constraints in the basis.  At each step, the basis is replaced by another one with a worse value of $\bar x$; when no more basis replacement steps are possible, the correct solution has been found.

Very quickly, workers in this area realized that similar techniques could also be applied to certain nonlinear programs such as the minimum enclosing ball problem~\cite{AdlSha-MP-93,Ame-DCG-94,ChaMat-SODA-93,Cla-JACM-95,Dye-SJC-84,Dye-SCG-92,FisGaeKut-ESA-03,Gar-SJC-95,Gae-ESA-99,MatShaWel-Algo-96,Meg-SJC-83,Pos-STOC-84,Wel-NRNTCS-91}.  One of the most popular and general formulations of this form of generalized linear program is the class of {\em LP-type problems}
defined by Matou\v{s}ek et al.~\cite{MatShaWel-Algo-96}; we follow the description of this formulation from Amenta et al.~\cite{AmeBerEpp-Algs-99}.

An LP-type problem consists of a finite set
$S$ of {\em constraints} and an {\em objective function} $f$ mapping
subsets of $S$ to some totally ordered space and satisfying the
following two properties:
\begin{enumerate}
\item For any $A\subset B$, $f(A)\le f(B)$.
\item For any $A$, $p$, and $q$,
if $f(A)=f(A\cup\{p\})=f(A\cup\{q\})$, then
$f(A)=f(A\cup\{p,q\})$.
\end{enumerate}
The problem is to compute
$f(S)$ using only evaluations of $f$ on small subsets of~$S$.

For instance, in linear programming, $S$ is a
set of halfspaces and $f(S)$ is the point in the intersection of the
halfspaces at which some linear function takes its minimum value. In the smallest enclosing ball problem, $S$ consists of the points
themselves, and $f(A)$ is the smallest enclosing ball of $A$, where the total ordering on balls is given by their radii.
It is not hard to see that this system satisfies the properties above: removing points can only make the radius shrink or stay the same, and if a ball contains the additional
points $p$ and $q$ separately it contains them both together.

A {\em basis} of an LP-type problem is a set $B$
such that for any $A\subsetneq B$, $f(A)<f(B)$.
Thus, due to the first property of an LP-type problem, the value of the overall problem is the
same as the value of the {\em optimal basis}, the basis $B$ that maximizes $f(B)$.
The {\em dimension} of an LP-type problem is the maximum
cardinality of any basis; although we have not included it above, a requirement that this dimension be bounded is often included in the definition of an LP-type problem.  The dimension of an LP-type problem
may differ from the dimension of some space $\R^d$ that may be associated in some way with the problem; for instance, for smallest enclosing balls in $\R^d$, the dimension of the LP-type problem
turns out to be $d+1$ instead of~$d$.

As Matou\v{s}ek et al.~\cite{MatShaWel-Algo-96} describe, efficient and simple randomized algorithms
for bounded-dimension LP-type problems are known, with running time
$O(dnT + t(d)E\log n)$ where $n$ is the number of
constraints,
$T$ measures the time to test whether $f(B)=f(B\cup\{x\})$ for some basis $B$ and element $x\in S$,
$t(d)$ is exponential or subexponential, and $E$ is the time to perform a {\em basis-change operation}
in which we must find the
basis of a constant-sized subproblem and use it to replace the current basis.
It is also possible with certain additional assumptions to solve these problems deterministically
in time linear in $n$~\cite{ChaMat-SODA-93}.

As Amenta et al.~\cite{AmeBerEpp-Algs-99} showed, quasiconvex programs can be expressed as LP-type problems, in such a way that the dimension of the LP-type problem is not much more than the dimension of the domain of the quasiconvex functions; therefore, quasiconvex programs can be solved in a linear number of function evaluations and a sublinear number of basis-change operations.

In order to specify the LP-type dimension of these problems, we need one additional definition:
suppose we have a nested convex family $\kappa_i$.  If $\kappa_i(\lambda)$ does not depend on $\lambda$, we say that $\kappa_i$ is {\em constant}; such constant families arose, for instance, in our treatment of convex programs.  Otherwise, suppose $\kappa_i$ is associated with a quasiconvex function $q_i$.
If there is no open set $S$ such that $q_i$ is constant over $S$,
and if $\kappa(t')$ is contained in the interior of
$\kappa(t)$ for any $t'<t$, we say that $\kappa$ is {\em continuously
shrinking}.  We note that this property is different from the related and more well-known property of {\em strict quasiconvexity} (a quasiconvex function is strictly quasiconvex if, whenever it is constant on a line segment, it remains constant along the whole line containing the segment): $L_1$ distance from the origin (in $\R^d$, $d>1$) is continuously shrinking but not strictly quasiconvex.  On the other hand, the function
$$f(x,y)=\min\{r\mid x^2+(y-r)^2\le r^2\}$$
(on the closed upper halfplane $y\ge 0$) is strictly quasiconvex but not continuously shrinking, since
the origin is on the boundary of all its level sets.

We repeat the analysis of Amenta et al.~\cite{AmeBerEpp-Algs-99}, showing that quasiconvex programs are LP-type problems, below.

\begin{theorem}\label{thm:lp-type}
Any quasiconvex program forms an LP-type problem of dimension at most $2d+1$.
If each $\kappa_i$ in the quasiconvex program is either
constant or continuously shrinking, the dimension is at most $d+1$.
\end{theorem}

\begin{proof}
We form an LP-type problem in which the set $S$ consists of the nested convex families defining the quasiconvex program, and the objective function $\Lambda(T)$ gives the value of the quasiconvex program defined by the nested convex families in $T$.
Then, 
property~1 of LP-type problems is obvious: adding another nested convex family to the input can only further constrain the solution values and increase the min-max solution.
To prove property~2, recall that $\Lambda(T)$ is defined as the minimum point
of the intersection $\{(\lambda,\bar x)\mid\bar x\in\kappa_i(\lambda)\}$
(the intersection is nonempty by the remark in Section~\ref{sec:qcp} about replacing infima by minima).
If this point belongs to the intersection for sets $A$, $A\cup\{\kappa_i\}$, and $A\cup\{\kappa_k\}$, then clearly it belongs to the intersection for $A\cup\{\kappa_i,\kappa_j\}$.
It remains only to show the stated bounds on the dimension.

First we prove the dimension bound for the general case, where we do not assume continuous
shrinking of the families in $S$. Let $(\lambda,\bar x)=\Lambda(S)$.
For any $\lambda'<\lambda$, 
$$\bigcap_{i\in S}\kappa_i(\lambda')=\emptyset,$$
so by Helly's theorem some $(d+1)$-tuple of sets $\kappa_i(\lambda')$ has empty
intersection.  If there is some $\lambda''<\lambda$ for which this $(d+1)$-tuple's intersection becomes nonempty, replace $\lambda'$ by $\lambda''$, find another $(d+1)$-tuple with empty intersection for the new $\lambda'$, and repeat until this replacement process terminates.  There are only finitely many possible $(d+1)$-tuples of nested convex families, and each replacement increases $\lambda'$, so the replacement process must terminate and we eventually find a $(d+1)$-tuple $B^{-}$ of nested convex families that has empty intersection for all $\lambda'<\lambda$.

With this choice of $B^{-}$, $\Lambda(B^{-})=(\lambda,\bar y)$ for some $\bar y$, so the presence of
$B^{-}$ forces the LP-type problem's solution to have the correct value of $\lambda$.
We must now add further nested convex families to our basis to force the solution to also have the correct value of $\bar x$.  Recall that
$$\bar x\in L=\bigcap_{i\in S}\kappa_i(\lambda),$$
and $\bar x$ is the minimal point in $L$.  By Helly's theorem again, the location of this minimal point is determined by some
$d$-tuple $B^{+}$ of the sets $\kappa_i(\lambda)$.  Then $\Lambda(B^{-}\cup
B^{+})=\Lambda(S)$, so some basis of $S$ is a subset of $B^{-}\cup B^{+}$ and
has cardinality at most $2d+1$.

Finally, we must prove the improved dimension bound for well-behaved nested convex families, so suppose  each $\kappa_i\in S$ is
constant or continuously shrinking.  Our strategy will be to again find a
tuple
$B^{-}$ that determines $\lambda$, and a tuple $B^{+}$ that
determines $\bar x$, but we will use continuity to make the sizes of
these two tuples add to at most $d+1$.

The set $L$ defined above has empty interior: otherwise, we could find an open region $X$
within $L$, and a nested family $\kappa_i\in S$ such that $\kappa_i(\lambda')\cap
X=\emptyset$ for any $\lambda'<\lambda$, violating the assumption that $\kappa_i$
is constant or continuously shrinking.
If the interior of some $\kappa_i(\lambda)$ contains a point of
the affine hull of $L$, we say that $\kappa_i$ is ``slack'';
otherwise we say that
$\kappa_i$ is ``tight''.  The boundary of a slack $\kappa_i(\lambda)$
intersects
$L$ in a subset of measure zero (relative to the affine hull of
$L$), so we can find a point
$\bar y$ in the relative interior of
$L$ and not on the boundary of any slack $\kappa_i$. Form the
projection
$\pi:\R^d\mapsto\R^{d-\dim L}$ onto the orthogonal complement of $L$.

For any ray $r$ in $\R^{d-\dim L}$ starting at the point
$\pi(L)$, we can lift that ray to a ray $\hat r$ in $\R^d$
starting at $\bar y$, and find a hyperplane
containing $L$ and separating the interior of some
$\kappa_i(\lambda)$ from $\hat r\setminus\{\bar y\}$.
This separated $\kappa_i$ must be
tight (because it has $\bar y$ on its boundary as the origin of the ray)
so the separating hyperplane must contain the affine hull of $L$
(otherwise some point in $L$ within a small neighborhood of $\bar x$
would be interior to $\kappa_i$).  Therefore the hyperplane
is projected by $\pi$ to a lower dimensional hyperplane separating
$\pi(\kappa_i(\lambda))$ from $\pi(L)$.
Since one can find such a separation for any ray,
$\bigcap_{{\rm tight}\,\kappa_i}\pi(\kappa_i(\lambda))$ can not contain any
points of any such ray and must consist of the single point
$\pi(L)$.
At least one tight $\kappa_j$
must
be continuously shrinking (rather than constant), since otherwise
$\bigcap_{\kappa_i\in S}\kappa_i(\lambda')$ would be nonempty for some $\lambda'<\lambda$.
The intersection of the interior of $\pi(\kappa_j(\lambda))$ with
the remaining projected tight constraints $\pi(\kappa_i(\lambda))$ is empty,
so by
Helly's theorem, we can find a $(d-\dim L + 1)$-tuple
$B^{-}$ of these convex sets having empty intersection, and the
presence of $B^{-}$ forces the LP-type problem's solution to have the correct value of
$\lambda$. Similarly, we can reduce the size of the set $B^{+}$ determining
$\bar x$ from $d$ to $\dim L$, so the total size of a basis
is at most $(d-\dim L + 1)+\dim L=d+1$.
\end{proof}

This result provides theoretically efficient combinatorial algorithms for quasiconvex programs, and allows us to claim $O(n)$ time randomized algorithms for most quasiconvex programming problems in the standard computational model for computational geometry, in which primitives of constant description complexity may be assumed to be solved in constant time.
For certain well-behaved sets of quasiconvex functions (essentially,
the family of sets $S_{\bar x,\lambda}=\{\kappa\in S\mid \bar x\in\kappa(\lambda)\}$ should
have bounded Vapnik-Chervonenkis dimension) the technique of Chazelle and Matousek~\cite{ChaMat-SODA-93} applies and these problems can be solved deterministically in $O(n)$ time.

However, we should note that there are some difficulties with this approach in practice.  In particular, although the basis-change operations have constant description complexity, it may not always be clear how to implement them efficiently.
Therefore, in the next section we discuss alternative numerical techniques for solving quasiconvex programs directly, based only on simpler operations (function and gradient evaluation).  It may be of interest to combine the two approaches, by using numerical techniques to solve the basis change operations needed for the LP-type approach; however, we do not have any theory describing how the LP-type algorithms might be affected by approximate numerical results in the basis-change steps.

\subsection{Implicit Quasiconvex Programming}

In some circumstances we may have a set of $n$ inputs that leads to a quasiconvex program with many more than $n$ quasiconvex functions; for instance, there may be one such function per pair of inputs.
If we directly apply an LP-type algorithm, we will end up with a running time much larger than the $O(n)$ input size.
Chan~\cite{Cha-SODA-04} showed that, in such circumstances, the time for solving the quasiconvex program can often be sped up to match the time for a {\em decision algorithm} that merely tests whether a given pair $(\lambda,\bar x)$ provides a feasible solution to the program.

As a simple example, consider a variation of the smallest enclosing ball problem.  Suppose that we wish to place a center that minimizes the maximum sum of distances to any $k$-tuple of sites, rather than (as in the smallest enclosing ball problem) minimizing the maximum distance to a single site.  This can be expressed again as a quasiconvex program: the sum of distances to any $k$-tuple of sites is quasiconvex, as it is a sum of convex functions.  There are $O(n^k)$ such functions, so the problem can be solved in $O(n^k)$ time by the methods discussed already.  However, the quality of any fixed center can easily be evaluated much more quickly, in $O(n)$ time, and Chan's technique provides an automatic method for turning this fast evaluation algorithm into a fast optimization algorithm for choosing the best center location.

Chan's result applies more generally to LP-type problems, but we state it here as it applies to implicit quasiconvex programming.

\begin{theorem}
\label{thm:iqcp}
Let $\cal Q$ be a space of quasiconvex functions,
$\cal P$ be a space of input values,
and $f:2^{\cal P}\mapsto 2^{\cal Q}$ map sets of input values to sets of functions in $\cal Q$.
Further, suppose that $\cal P$, $f$, and $\cal S$ satisfy the following properties:
\begin{itemize}
\item There exists a constant-time subroutine for solving quasiconvex programs
of the form $f(B)$ for any $B\subset\cal P$ with $|B|=O(1)$.
\item There exists a decision algorithm
that takes as input a set $P\subset\cal P$ and a pair $(\lambda,\bar x)$, and returns yes
if and only if $\bar x\in\kappa(\lambda)$ for all $\kappa\in f(P)$.
The running time of the decision algorithm is bounded by $D(|P|)$, where
there exists a constant $\epsilon>0$ such that $D(n)/n^{\epsilon}$ is monotone increasing.
\item There are constants $\alpha$ and $r$ such that, for any input set $P\subset\cal P$, we can find in time at most $D(|P|)$ a collection of sets $P_i$,
$0\le i<r$, each of size at most $\alpha |P|$, for which
$f(P)=\cup_i f(P_i)$.
\end{itemize}
Then for any $P\subset\cal P$ we can solve the quasiconvex program
$f(P)$, where $|P|=n$, in randomized expected time $O(D(n))$.
\end{theorem}

The proof involves solving a slightly more general problem in which
we are given, not just a single input $P$, but a set of inputs $P_1$, $\ldots$, $P_d$,
where $d$ is the dimension of the LP-type problems coming from $\cal Q$,
and must solve the quasiconvex program $\cup f(P_i)$.
Given any such problem, we partition each input $P_i$ into $r^i$ subproblems
$P_{i,j}$ of size at most $\alpha^i n$ for an appropriately chosen $i$, by repeatedly subdividing large subproblems into smaller ones.
We then view the subproblems $P_{i,j}$ as being constraints for an LP-type problem
in which the objective
function is the solution to the quasiconvex program
$\cup_{P_{i,j}\in S} f(P_{i,j})$.
This new LP-type problem turns out to have the same dimension as the quasiconvex programs with which we started, and the result follows by applying a standard LP-type algorithm to 
this problem and solving the divide-and-conquer recurrence that results.

The first and last conditions of the theorem are easily met when $f(P)$ produces one or a constant number of quasiconvex functions per $k$-tuple of inputs for some constant $k$ (as in our example of optimizing the sum of $k$ distances): then,
constant sized input sets lead to constant sized quasiconvex programs, and
if the input is partitioned into $k+1$ equal-sized subsets, the complements of these
subsets provide the sets $P_i$ needed for the last condition.
For such problems, the main difficulty in applying this theorem is finding an appropriate decision algorithm.
For our example of minimizing the maximum sum of $k$ distances, the decision algorithm is also straightforward (select and add the $k$ largest distances from the given center to the sites) and so we can apply Chan's result to solve this problem in $O(n)$ time.

Chan's implicit quasiconvex programming algorithm is important in the robust statistics application described later.  This algorithm has also been applied to problems of inverse parametric minimum spanning tree computation~\cite{Cha-SODA-04,Epp-SJC-03} and facility location~\cite{ew-arxiv}.

\subsection{Smooth Quasiconvex Programming}

\begin{figure*}[t]
\centering\includegraphics[width=4in]{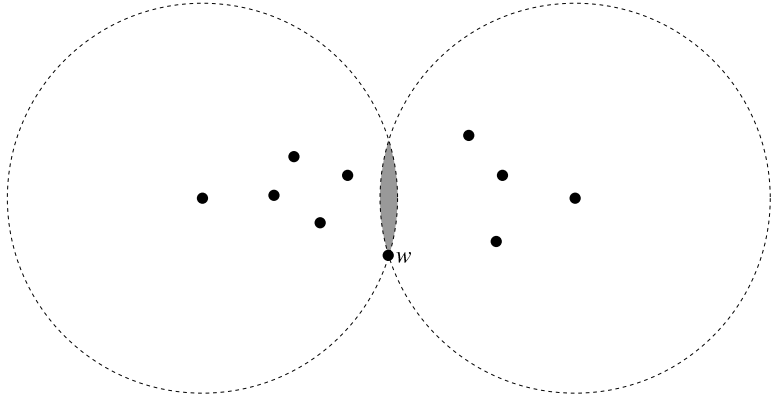}
\caption{Example showing the difficulty of applying standard gradient descent methods to quasiconvex programming.  The function to be minimized is the maximum distance to any point;
only points within the narrow shaded intersection of circles have function values smaller than the value at point~$w$.  Figure taken from~\cite{Epp-SODA-04-qaba}.}
\label{fig:cr}
\end{figure*}

If all functions $q_i(\bar x)$ are quasiconvex, the function $q(\bar x)=\max_i q_i(\bar x)$ is itself quasiconvex, so we can apply hill-climbing procedures to find its infimum.  Such hill climbing procedures may be desirable in preference to the combinatorial algorithms for LP-type problems, as they avoid the difficulty of describing and implementing an appropriate exact basis change procedure.  In addition, a hill climbing information that uses only numerical evaluation of function values (or possibly also function gradient evaluations) can be implemented in a generic way that does not depend on the specific form of the quasiconvex functions given to it as input.

However, many of the known non-linear optimization techniques require the function being optimized to satisfy some smoothness conditions.  In many of our applications the individual functions $q_i$ are smooth, but their maximum $q$ may not be smooth, so it is difficult to apply standard gradient descent techniques.  The difficulty may be seen, for instance, in the smallest enclosing ball problem in the plane (Figure~\ref{fig:cr}).  A basis for this problem may consist of either two or three points.  If a point set has only two points in its basis, and our hill climbing procedure for circumradius has reached a point $w$ equidistant from these two points and near but not on their midpoint, then improvements to the function value $q(w)$ may be found only by moving $w$ in a narrow range of directions towards the midpoint.  Standard gradient descent algorithms may have a difficult time finding such an improvement direction.  

To avoid these difficulties, we introduced in~\cite{Epp-SODA-04-qaba} the following algorithm, which we call {\em smooth quasiconvex programming}, and which can be viewed as a generalization of Zoutendijk's method of feasible directions~\cite{Zou-60} for convex programming.  
If a quasiconvex function $q_i$ is differentiable, and $w$ is a point where $q_i$ is not minimal,
then one can find a point with a smaller value by moving a sufficiently small distance from $x$ along any direction having negative dot product with the gradient of $q_i$ at~$w$.  Thus, we can improve $q(w)$ by moving in a direction that is negative with respect to all the gradients of the functions that determine the value of $q(w)$.

We formalize this notion and generalize it to nondifferentiable functions as follows.  Assume for the purposes of this algorithm that, for each of the input quasiconvex functions $q_i$, and each $\bar x$ that is not the minimum point of $q_i$, we also can compute a vector-valued function $q_i^*(\bar x)$,
satisfying the following properties:
\begin{enumerate}
\item If $q_i(\bar y)<q_i(\bar x)$, then $(\bar y - \bar x)\cdot q_i^*(\bar x)>0$, and
\item If $q_i^*(\bar x)\cdot \bar y>0$, then for all sufficiently small $\epsilon>0$, $q_i(\bar x + \epsilon\bar y)< q_i(\bar x)$.
\end{enumerate}
Less formally, any vector $\bar y$ is an improving direction for $q_i(\bar x)$ if and only if it has positive inner product with $q_i^*(\bar x)$.

If the level set $q_i^{\le\lambda}$
is a {\em smooth} convex set (one that has at each of its boundary points a unique tangent plane),
then the vector $q_i^*(\bar x)$ should be an inward-pointing normal vector to the tangent plane
to $q_i^{\le q(\bar x)}$ at $\bar x$.
For example, in the smallest enclosing ball problem, the level sets are spheres, having tangent planes perpendicular to the radii, and $q_i^*$ should point inwards along the radii of these spheres.
If $q_i$ is differentiable then $q_i^*$ can be computed as the negation of the gradient of $q_i$,
but the functions $q_i^*$ also exist for discontinuous functions with smooth level sets.

Our smooth quasiconvex programming algorithm begins by selecting an initial value for $\bar x$, and a desired output tolerance.
Once these values are selected, we repeat the following steps:
\begin{enumerate}
\item Compute the set of vectors $q_i^*(\bar x)$,
for each $i$ such that $q_i(\bar x)$ is within the desired tolerance of $\max_i q_i(\bar x)$.
\item Find an improving direction $\bar y$; that is, a vector such that $\bar y\cdot q_i^*(\bar x)>0$ for each vector $q_i^*(\bar x)$ in the computed set.
If no such vector exists, $q(\bar x)$ is within the tolerance of its optimal value and the algorithm terminates.
\item Search for a value $\epsilon$ for which $q(\bar x+\epsilon\bar y)\le q(\bar w)$,
and replace $\bar x$ by $\bar x+\epsilon\bar y$.
\end{enumerate}

The search for a vector $\bar y$ in step 2 can be expressed as a linear program.
However, when the dimension of the quasiconvex functions' domain is at most two
(as in the planar smallest enclosing ball problem)
it can be solved more simply by sorting the vectors $q_i^*(\bar x)$ radially around the origin
and choosing $\bar y$ to be the average of two extreme vectors.

In step 3, it is important to choose $\epsilon$ carefully.
It would be natural, for instance, to choose $\epsilon$ as large as possible while satisfying the inequality in that step; such a value could be found by a simple doubling search.  However, such a choice could lead to situations where the position of $\bar x$ oscillates back and forth across the true optimal location.  Instead, it may be appropriate to reduce the resulting $\epsilon$ by a factor of two before replacing~$\bar x$.

We do not have any theory regarding the convergence rate of the smooth quasiconvex programming algorithm, but we implemented it and applied it successfully in the automated algorithm analysis application discussed below~\cite{Epp-SODA-04-qaba}.  Our implementation appeared to exhibit linear convergence: each iteration increased the number of bits of precision of the solution by a constant.
Among numerical algorithms techniques, the sort of gradient descent we perform here is considered naive and inefficient compared to other techniques such as conjugate gradients or Newton iteration, and it would be of interest to see how well these more sophisticated methods could be applied to quasiconvex programming.

\section{Applications}

We have already described some simple instances of geometric optimization problems that can be formulated as quasiconvex programs.  Here we describe some more complex applications of geometric optimization, in which quasiconvex programming plays a key role.

\subsection{Mesh Smoothing}
\label{sec:meshsmooth}

An important step in many scientific computation problems, in which differential equations describing airflow, heat transport, stress, global illumination, or similar quantities are simulated, is {\em mesh generation}~\cite{BerEpp-CEG-95,BerPla-HCG-00}.  In this step, a complex two- or three-dimensional domain is partitioned into simpler regions, called elements, such as triangles or quadrilaterals in the plane or tetrahedra or cuboids in three dimensions.  Once these elements are formed, one can then set up simple equations relating the values of the quantity of interest in each of the elements, and solve the equations to produce the results of the simulation.  In this section we are particularly concerned with {\em unstructured mesh generation}, in which the pattern of connections from element to element does not form a regular grid; we will consider a problem in structured mesh generation in a later section.

In meshing problems, it is important to find a mesh that has small elements in regions of fine detail, but larger elements elsewhere, so that the total number of elements is minimized; this allows the system of equations derived from the mesh to be solved quickly.  It is also important for the accuracy of the simulation that the mesh elements be {\em well shaped}; typically this means that no element should have very sharp angles or angles very close to $180^\circ$.  To achieve a high quality mesh,
it is important not only to find a good initial placement of mesh vertices (the main focus of most meshing papers) but then to modify the mesh by changing its topology and moving vertices until no further quality increase can be achieved.  We here concentrate on the problem of moving mesh vertices while retaining a fixed mesh topology, known as {\em mesh smoothing}~\cite{AmeBerEpp-Algs-99,BanSmi-SJNA-97,CanTriSta-IMR-98,Dji-IMR-00,Fre-McNU-97,FreJonPla-IMR-95,FreJonPla-SJSC-99,FreOll-IJNME-97,VolMenMol-EG-99}.

\begin{figure}[t]
\centering
\includegraphics[width=2in]{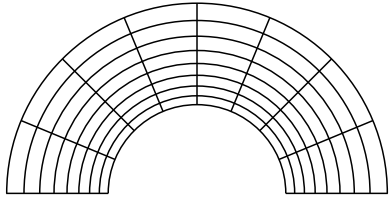}
\caption{Mesh of an arched domain. Too much Laplacian smoothing can lead to invalid placements of the internal vertices beyond the boundaries of the arch.}
\label{fig:Bridge}
\end{figure}

\begin{figure}[t]
\centering
\includegraphics[width=2in]{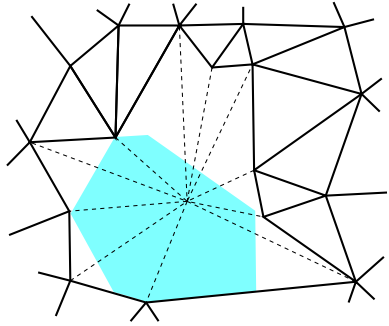}
\caption{Optimization-based smoothing of a triangular mesh in $\R^2$. At each step we remove a vertex from the mesh, leaving a star-shaped polygon, then add a new vertex within the kernel (shaded) of the star-shaped region and retriangulate.  Figure taken from~\cite{AmeBerEpp-Algs-99}.}
\label{fig:kernel}
\end{figure}

\begin{figure}[t]
\centering
\includegraphics[width=4in]{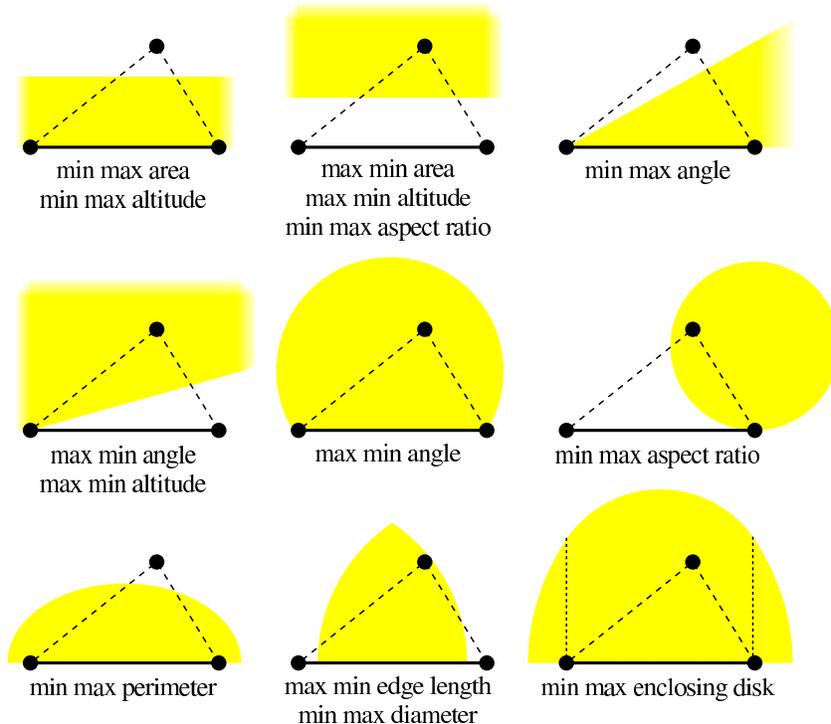}
\caption{Level set shapes for various mesh element quality measures.  Figure modified from one in~\cite{AmeBerEpp-Algs-99}.}
\label{fig:MeshSmooth}
\end{figure}

Two approaches to mesh smoothing have commonly been used, although they may sometimes be combined~\cite{CanTriSta-IMR-98,Fre-McNU-97}:
In {\em Laplacian smoothing}, all vertices are moved towards the centroid of their neighbors. Although this is easy and works well for many instances, it has some problems; for instance in a regular mesh on an arched domain (Figure~\ref{fig:Bridge}), repeated Laplacian smoothing can cause the vertices at the top of the arch to sag downwards, eventually moving them to invalid positions beyond the boundaries of the domain.

Instead, {\em optimization-based smoothing} takes a more principled approach, in which we decide on a measure of element quality that best fits our application, and then seek the vertex placement that optimizes that quality measure.  However, since simultaneous global optimization of all vertex positions seems a very difficult problem, we instead cycle through the vertices optimizing their positions a single vertex at a time.  At each step (Figure~\ref{fig:kernel}), we select a vertex and remove it from the mesh, leaving a star-shaped region consisting of the elements incident to that vertex.  Then, we place a new vertex within the kernel of the star-shaped region, and form a mesh again by connecting the new vertex to the boundary of the region.  Each step improves the overall mesh quality, so this optimization process eventually converges to a locally optimal placement, but we have no guarantees about its quality with respect to the globally optimal placement.

However, in the individual vertex placement steps we need accept no such compromises with respect to global optimization.  As we showed in~\cite{AmeBerEpp-Algs-99}, for many natural measures $q_i(\bar x)$ of the quality of an element incident to vertex $\bar x$ (with smaller numbers indicating better quality), the problem of finding a mesh minimizing the maximum value of $q_i$ can be expressed as a quasiconvex program.  Figure~\ref{fig:MeshSmooth} illustrates the level set shapes resulting from various of these quasiconvex optimization-based mesh smoothing problems.
For shape-based quality measures, such as maximizing the minimum angle, the optimal vertex placement will naturally land in the interior of the kernel of the region formed by the removal of the previous vertex placement.  For some other quality measures, such as minimizing the maximum perimeter, it may be appropriate to also include constant quasiconvex functions, forcing the vertex to stay within the kernel, similar to the functions used in our transformation of convex programs to quasiconvex programs.  It would also be possible to handle multiple quality measures simultaneously by including quasiconvex functions of more than one type in the optimization problem.

In most of the cases illustrated in Figure~\ref{fig:MeshSmooth}, it is straightforward to verify that the quality measure has level sets of the convex shape illustrated.  One possible exception is the problem of minimizing the maximum aspect ratio (ratio of the longest side length to shortest altitude) of any element.  To see that this forms a quasiconvex optimization problem, Amenta et al.~\cite{AmeBerEpp-Algs-99} consider separately the ratios of the
three sides to their corresponding altitudes; the maximum of these three
will give the overall aspect ratio.  The ratio of a side external to the star to its corresponding
altitude has a feasible region (after taking into account the
kernel constraints) forming a halfspace parallel to the external side,
as shown in Figure~\ref{fig:MeshSmooth} (top center).
To determine the aspect ratio on
one of the other two sides of a triangle $\Delta_i$, normalize the
triangle coordinates so that the replaced point has coordinates $(x,y)$ and
the other two have coordinates $(0,0)$ and $(1,0)$.  The side length is
then $\sqrt{x^2+y^2}$, and the altitude is $y/\sqrt{x^2+y^2}$,
so the overall aspect ratio has the simple formula $(x^2+y^2)/y$.
The locus of points for which this is a constant $b$
is given by $x^2+y^2=by$, or equivalently
$x^2+(y-(b/2))^2=(b/2)^2$.  Thus the feasible region is a
circle tangent to the fixed side of $\Delta_i$ at one of its two
endpoints (Figure~\ref{fig:MeshSmooth}, center right).
Another nontrivial case is that of minimizing the smallest enclosing ball of the element,
shown in the bottom right of the figure; in that case the level set boundary consists of curves of two types, according to whether, for placements in that part of the level set, the enclosing ball touches two or three of the element vertices, but the curves meet at a common tangent point to form a smooth convex level set. 

Bank and Smith~\cite{BanSmi-SJNA-97} define yet another measure of the quality of a
triangle, computed by dividing the triangle's area by the sum of the
squares of its edge lengths.  This gives a dimensionless quantity which
Bank and Smith normalize to be one for the equilateral triangle (and
less than one for any other triangle).
As Bank and Smith show, the lower level sets for this mesh quality measure
form circles centered on the perpendicular bisector of the two fixed
points of the mesh element, so, as with the other measures, finding the placement
optimizing Bank and Smith's measure can be expressed as a quasiconvex program. 

We have primarily discussed triangular mesh smoothing here, but the same techniques
apply with little modification to many natural element quality measures for quadrilateral and tetrahedral mesh smoothing.
Smoothing of cubical meshes is more problematic, though, as moving a single vertex may
cause the faces of one of the cuboid elements to become significantly warped.
Several individual quasiconvex quality measures for quadrilateral and tetrahedral meshes, and the shapes of their level sets, are discussed in more detail in~\cite{AmeBerEpp-Algs-99}.
The most interesting of these from the mathematical viewpoint is the problem of
maximizing the minimum solid angle of any tetrahedral element, as measured at its vertices,
which with some difficulty we were able to show leads to a quasiconvex objective function.

\subsection{Graph Drawing}

\begin{figure}[t]
\centering
\includegraphics[width=1.5in]{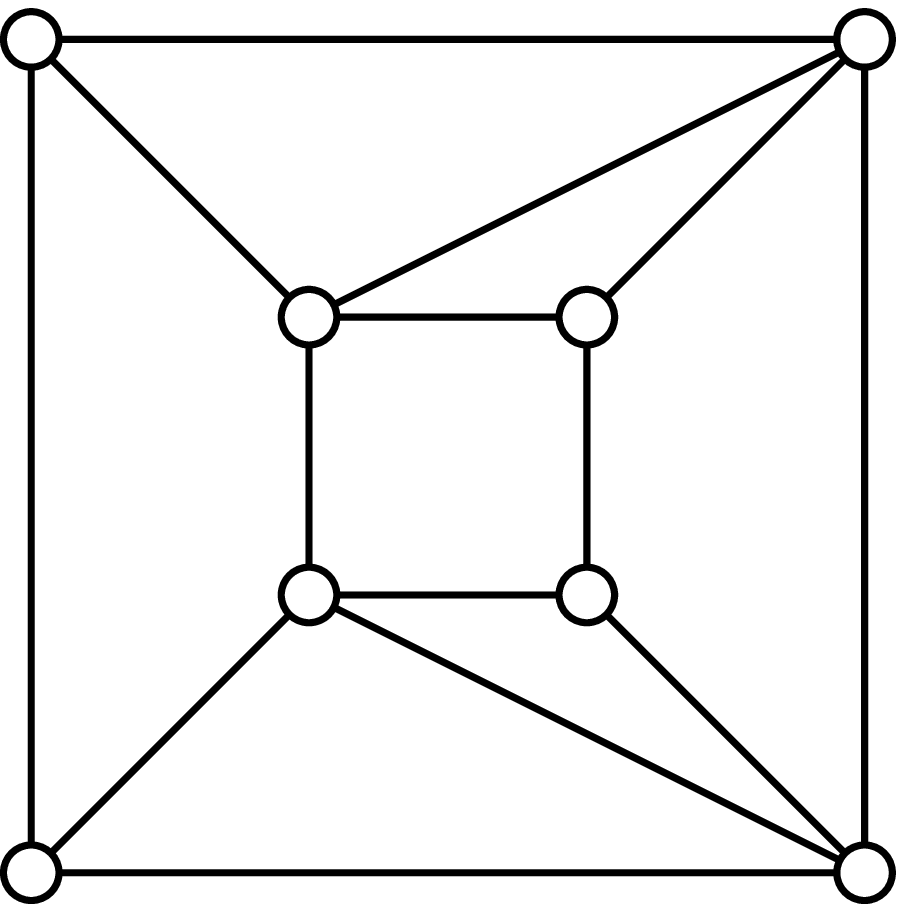}
\qquad
\includegraphics[width=1.5in]{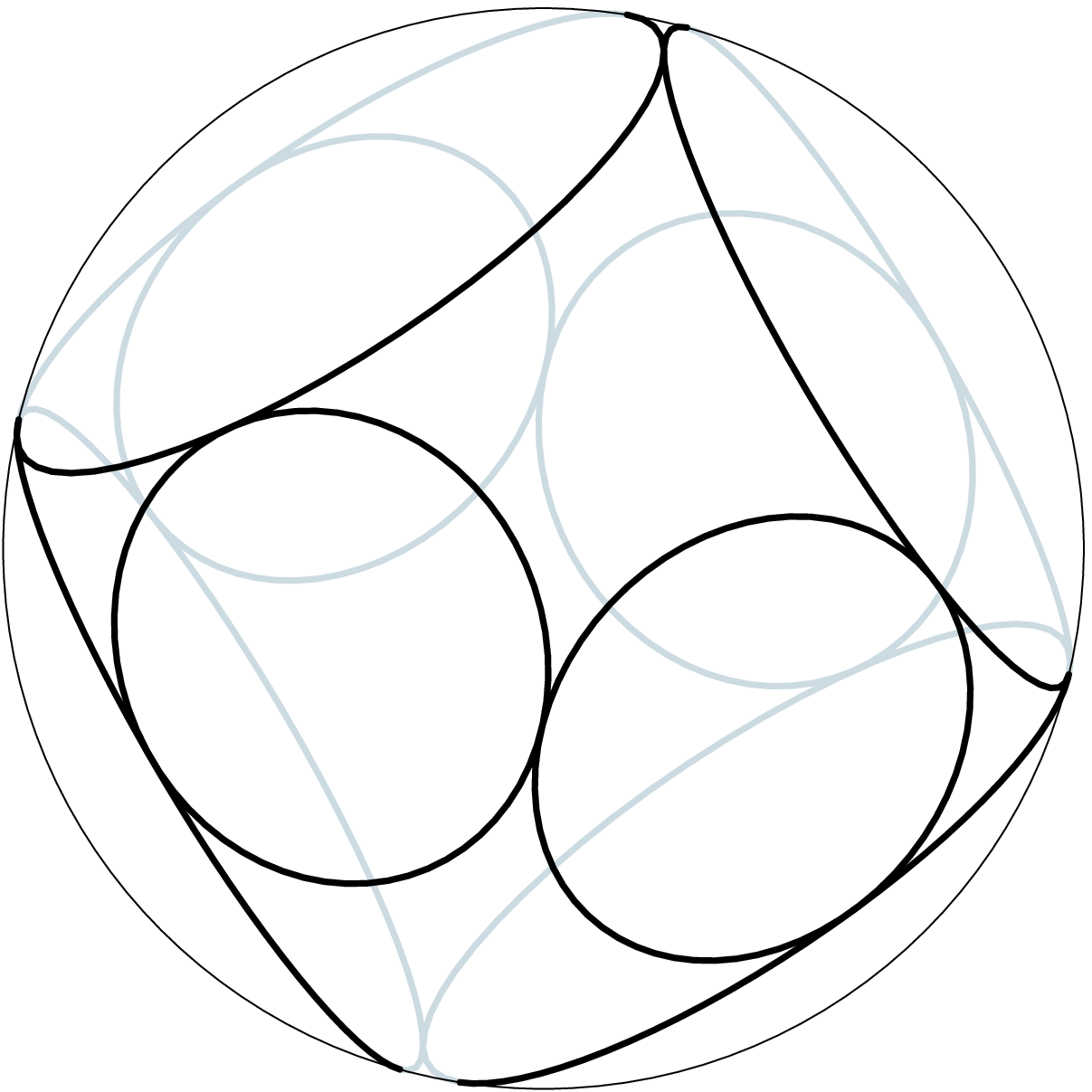}
\qquad
\includegraphics[width=1.5in]{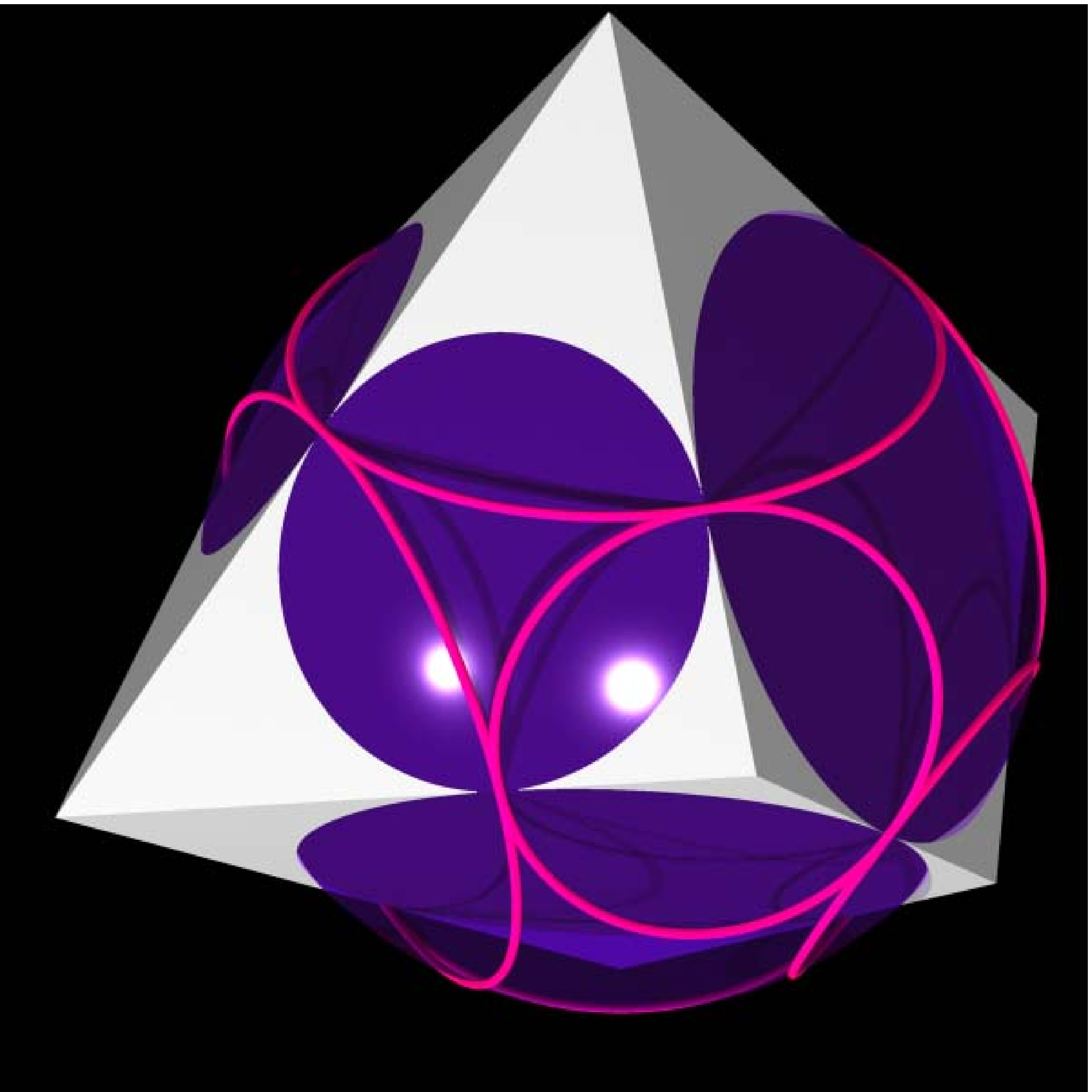}
\caption{Planar graph (left), its representation as a set of tangent disks on a sphere (center),
and the corresponding polyhedral representation (right). Left and center images taken from~\cite{BerEpp-WADS-01-omt}.}
\label{fig:Koebe}
\end{figure}

The Koebe-Thurston-Andreev embedding theorem~\cite{BriSch-SJDM-93,Koe-BSAW-36,Sac-DM-94}
states that any planar graph embedding can be transformed into a collection
of disks with disjoint interiors on the surface of a sphere, one disk per vertex,
such that two disks are tangent if and only if the corresponding two vertices are adjacent (Figure~\ref{fig:Koebe}, left and center).  The representation of the graph as such a collection of tangent disks is sometimes called a {\em coin graph}.
For maximal planar graphs, this coin graph representation is unique up to {\em M\"obius transformations}
(the family of transformations of the sphere that transform circles to circles),
and for non-maximal graphs it can be made unique by adding a new vertex within each face of the embedding,
adjacent to all vertices of the face, and finding a disk representation of the resulting augmented maximal planar graph.

Given a coin graph representation,
the graph itself can be drawn on the sphere e.g. by placing a vertex at the center of each circle and connecting two vertices by edges along an arc of a great circle; similar drawings are also possible in the plane by using polar projection to map the circles in the sphere onto circles in the plane~\cite{Hli-GD-97}.  Coin graphs can also be used to form a three-dimensional {\em polyhedral representation} of the graph, as follows: embed the sphere in space, and, for each disk, form a cone in space that is tangent to the sphere at the disk's boundary; then, form a polyhedron by taking the convex hull of the cone apexes.  The resulting polyhedron's skeleton is isomorphic to the original graph, and its edges are tangent to the sphere (Figure~\ref{fig:Koebe}, right).

In order to use these techniques for visualizing graphs, we would like to choose a coin graph representation that leads to several desirable properties identified as standard within the graph drawing literature~\cite{diBEadTam-99}, including the display of as many as possible of the symmetries of the original graph, and the separation of vertices as far apart from each other as possible. Our paper with Bern~\cite{BerEpp-WADS-01-omt} used quasiconvex programming to formalize the search for a drawing based on these objectives.

In order to understand this formalization, we need some more background knowledge about M\"obius transformations and their relation to hyperbolic geometry.  We can identify the unit sphere that the M\"obius transformations transform as being the boundary of a Poincar\'e or Klein model of hyperbolic space $\H^3$.  The points on the sphere can be viewed as ``infinite'' points that do not belong to $\H^3$ but are the limit points of certain sequences of points within $\H^3$.  With this identification, circles on the sphere become the limit points of hyperplanes in $\H^3$.  Any isometry of $\H^3$ takes hyperplanes to hyperplanes, and therefore can be extended to a transformation of the sphere that takes circles to circles, and the converse turns out to be true as well.  We can determine an isometry of $\H^3$ by specifying which point of $\H^3$ is mapped to the center of the Poincar\'e or Klein model, and then by specifying a spatial rotation around that center point.  The rotation component of this isometry does not change the shape of objects on the sphere, so whenever we seek the M\"obius transformation that optimizes some quality measure of a transformed configuration of disks on the sphere, we can view the problem more simply as one of seeking the optimal center point of the corresponding isometry in $\H^3$.

\begin{figure}[t]
\centering
\includegraphics[width=2in]{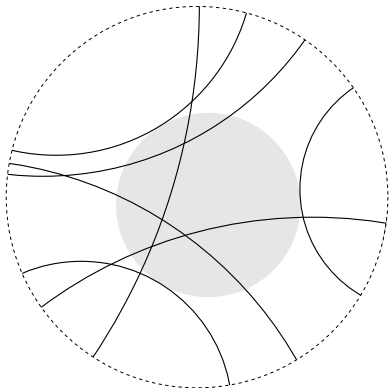}
\caption{Two-dimensional analogue of max-min radius transform problem: find the smallest disk touching all of a collection of hyperbolic lines.}
\label{fig:hbtouchline}
\end{figure}

To see how we apply this technique to our graph drawing problem, first consider a version of the problem in which we seek a disk representation maximizing the radius of the smallest disk.  More generally, given any collection of circles on the sphere, we wish to transform the circles in order to maximize the minimum radius.  Thus, let $q_i(\bar x)$ measure the (negation of the) transformed radius of the $i$th circle, as a function of the transformed center point $\bar x\in\H^3$.  If we let $H_i$ denote the hyperplane in $\H^3$ that has the $i$th circle as its set of limit points, then the transformed radius is maximized when the circle is transformed into a great circle; that is, when $\bar x\in H_i$.   If we choose a center point $\bar x$ away from $H_i$, the transformed radius will be smaller, and due to the uniform nature of hyperbolic space the radius can be written as a function only of the distance from $\bar x$ to $H_i$, not depending in any other way on the location of $\bar x$.  That is, the level sets of $q_i$ are the convex hyperbolic sets within some distance $R$ of the hyperplane $H_i$.
Therefore, $q_i$ is a quasiconvex hyperbolic function.  In fact, the quasiconvex program defined by
the functions $q_i$ can be viewed as a hyperbolic version of a generalized minimum enclosing ball problem, in which we seek the center $\bar x$ of the smallest ball that touches each of the convex sets $H_i$.  The two-dimensional version of this problem, in which we seek the smallest disk touching each of a collection of hyperbolic lines, is illustrated in Figure~\ref{fig:hbtouchline}.  If we form a Klein or Poincar\'e model with the resulting optimal point $\bar x$ at the center of the model, the corresponding M\"obius transformation of the model's boundary maximizes the minimum radius of our collection of circles.

Further, due to the uniqueness of quasiconvex program optima,
the resulting disk representation must display all the symmetries possible for the original planar graph embedding; for, if not all symmetries were displayed, one could use an undisplayed symmetry to relabel the vertices of the disk representation, achieving a second disk representation with equal quality to the first.  For instance, in Figure~\ref{fig:Koebe}, the disk representation shown has three planes of mirror symmetry while the initial drawing has only one mirror symmetry axis.

Bern and Eppstein~\cite{BerEpp-WADS-01-omt} then consider an alternative version of the graph drawing problem, in which the objective is to maximize the minimum distance between certain pairs of vertices on the sphere surface.  For instance, one could consider only pairs of vertices that are adjacent in the graph, or instead consider all pairs; in the latter case we can reduce the number of pairs that need be examined by the algorithm by using the Delaunay triangulation in place of the complete graph.
The problem of maximizing the minimum spherical distance among a set of pairs of vertices can be formulated as a quasiconvex program by viewing each pair of vertices as the two limit points of a hyperbolic line in $\H^3$, finding the center $\bar x$ of the smallest ball in $\H^3$ that touches each of these hyperbolic lines, and using this choice of center point to transform the sphere.

M\"obius transformations can also be performed on the augmented plane $\R^2\cup\{\infty\}$ instead of on a sphere, and act on lines and circles within that plane; a line can be viewed as a limiting case of a circle that passes through the special point $\infty$. 
Multiplication of each coordinate of each point by the same constant $k$ forms a special type of M\"obius transformation, which (if $k>1$) increases every distance, so it does not make sense to look for an unrestricted M\"obius transformation of the plane that maximizes the minimum Euclidean distance among a collection of pairs of points.  However, Bern and Eppstein were able to show, given a collection of points within the unit ball in the plane, that seeking the M\"obius transformation that takes that disk to itself and maximizes the minimum distances between certain pairs of the points can again be expressed as a two-dimensional quasiconvex program.  The proof of quasiconvexity is more complex and involves simultaneously treating the unit ball as a Poincar\'e model of $\H^2$ and the entire plane as the boundary of a Poincar\'e model of $\H^3$.

Along with these coin graph based drawing methods, Bern and Eppstein also considered a different graph drawing question, more directly involving hyperbolic geometry.
The Poincar\'e and Klein models of projective geometry have been considered by several authors as a way of achieving a ``fisheye'' view of a large graph, so that a local neighborhood in the graph is visible in detail near the center of the view while the whole graph is spread out on a much smaller scale at the periphery~\cite{LamRaoPir-CHI-95,Mun-CGA-97,MunBur-VRML-95}.
Bern and Eppstein~\cite{BerEpp-WADS-01-omt} found quasiconvex programming formulations of several versions of the problem of selecting an initial viewpoint for these hyperbolic drawings, in order for the whole graph to be visible in as large a scale as possible. For instance, a natural version of this problem would be to choose a viewpoint minimizing the maximum hyperbolic distance to any vertex, which is just the hyperbolic smallest enclosing ball problem again.  One question in this area that they left open is whether one can use quasiconvex programming to find a Klein model of a given graph that maximizes the minimum
Euclidean distance between adjacent vertices.

\subsection{Conformal Mesh Generation}

\begin{figure}[t]
\centering
\includegraphics[width=4in]{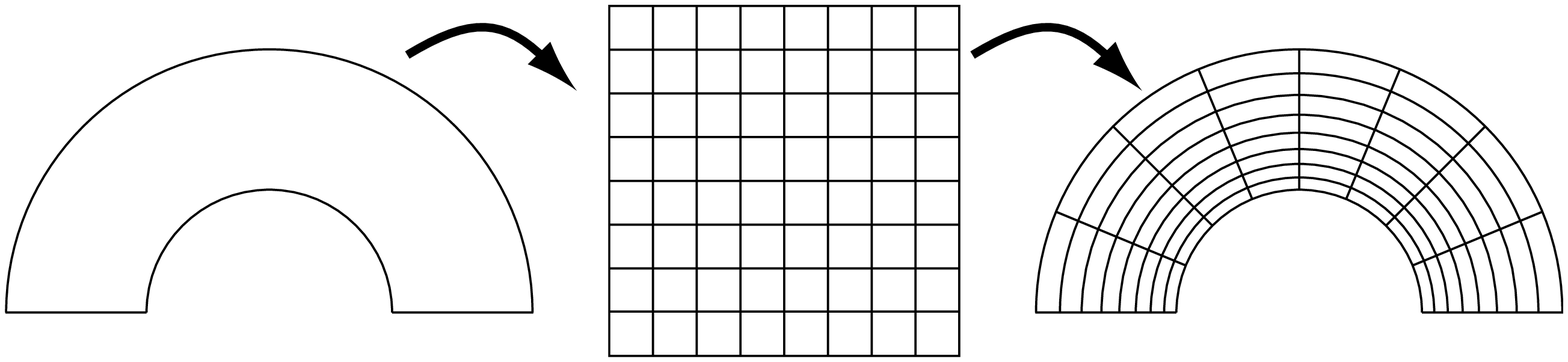}
\caption{Conformal meshing: transform domain to a more simply shaped region with a known mesh,
then invert the transformation to transform the mesh back to the original domain.}
\label{fig:confmesh}
\end{figure}

The ideas of mesh generation and optimal M\"obius transformation coincide in the problem of conformal mesh generation~\cite{BerEpp-WADS-01-omt}.  In this problem, we wish to generate a mesh for a simply-connected domain in $\R^2$ by using a {\em conformal transformation} (that is, a transformation that preserves angles of incidence between transformed curves) to map the shape into some easy-to-mesh domain such as a square, then invert the transformation to map the meshed square back into the original domain (Figure~\ref{fig:confmesh}).  There has been much work on algorithms for finding conformal
maps~\cite{DriVav-SISC-98,How-PhD-90,Smi-91,SteSch-CMFT-97,Tre-SSC-80} and conformal meshes have significant
advantages: the orthogonality of the angles at mesh vertices means that one can avoid
certain additional terms in the definition of the partial differential
equation to be solved~\cite{BerPla-HCG-00,ThoWarMas-85}.

If we replace the square in Figure~\ref{fig:confmesh} by a disk, the Riemann mapping theorem tells us that a conformal transformation always exists and is, moreover, unique up to M\"obius transformations that transform the disk to itself; any such transformation preserves conformality.
Thus, we have several degrees of freedom for controlling the size of the mesh elements produced by the conformal method: we can use a larger or smaller grid on the disk or square, but we can also use a M\"obius transformation in order to enlarge certain portions of the domain and shrink others before meshing it.  We would like to use these degrees of freedom to construct a mesh that has small elements in regions of the domain where fine detail is desired, and large elements elsewhere, in order to limit the total number of elements of the resulting mesh.

Bern and Eppstein~\cite{BerEpp-WADS-01-omt} formalized the problem by assuming an input domain in which certain
interior points $p_i$ are marked with a desired element size $s_i$.
If we find a conformal map $f$ from the domain to a disk,
the gradient of $f$ maps the marked element sizes
to desired sizes $s'_i$ in the transformed disk:
$s'_i = || f' (p_i) ||$.
We can then choose a structured mesh with element size $\min s'_i$
in the disk, and transform it back to a mesh of the original domain.
The goal is to choose our conformal map in a way that maximizes $\min
s'_i$, so that we can use a structured mesh with as few elements as
possible.  Another way of interpreting this is that
$s'_i$ can be seen as the radius of a small disk
at $f(p_i)$.  What we seek is the transformation that maximizes the minimum
of these radii.  This is not quite the same as the max-min radius graph drawing problem
of the previous section, because the circles to be optimized belong to $\R^2$ instead of to a sphere,
but as in the previous section we can view the unit disk as being a Poincar\'e model of $\H^2$
(using the fact that circles in $\H^2$ are mapped by the Poincar\'e model into circles in the unit disk),
and seek a hyperbolic isometry that maps $\H^2$ into itself and optimizes the circle radii.
The transformed radius of a circle is a function only of the distance from that circle
to the center point of the transformed model, so the level sets of the functions representing the transformed radii are themselves circles and the functions are quasiconvex.

The quasiconvex conformal meshing technique of Bern and Eppstein does not account for two remaining degrees of freedom: first, it is possible to rotate the unit disk around its center point and, while that will not change the element size as measured by Bern and Eppstein's formalization, it will change the element orientations.  This is more important if we also consider the second degree of freedom, which is that instead of using a uniform grid on a square, we could use a rectangle with arbitrary aspect ratio.  Bern and Eppstein leave as an open question whether we can efficiently compute the optimal choice of conformal map to a high-aspect-ratio rectangle to maximize the minimum desired element size.

\subsection{Brain Flat Mapping}

Hurdal et al.~\cite{HurBowSte-TR-99} describe methods for visualizing the complicated structure of the
brain by stretching its surface onto a flat plane.  
They perform this stretching via
conformal maps: surfaces of major brain components such as
the cerebellum are simply connected, so there exists a conformal map from these surfaces onto a
Euclidean unit disk, sphere, or hyperbolic plane.  Hurdal et al.{}
approximate this conformal map by using a fine triangular mesh to
represent the brain surface, and forming the Koebe disk representation of
this mesh.  Each triangle from the brain surface can then be mapped to
the triangle connecting the corresponding three disk centers.
As in the conformal meshing example, there is freedom to modify the conformal map by means of a M\"obius transformation, so
Bern and Eppstein~\cite{BerEpp-WADS-01-omt} suggested that the optimal M\"obius transformation technique described in the previous two sections could also be useful in this application,

Although conformal transformation preserves angles, it distorts other important geometric information such as area.  Bern and Eppstein proposed to ameliorate this distortion by using an optimal M\"obius transformation to find the conformal transformation minimizing the maximum ratio $a/a'$ where $a$ is the
area of a triangle in the initial three-dimensional map, and $a'$ is the
area of its image in the flat map.

Unfortunately it has not yet been possible to prove that this optimization problem leads to quasiconvex optimization problems.  Bern and Eppstein formalized the difficulty in the following open question:
Let $T$ be a triangle in the unit disk or on the surface of a sphere, and let $C$ be the set
of center points for Poincar\'e models (of $\H^2$ in the disk case or $\H^3$ in the sphere case)
such that the M\"obius transformations corresponding to center points in $C$
transform $T$ into a triangle of area at least $A$.  Is $C$ necessarily convex?
Note that, at least in the spherical case, the area of the transformed triangle is the same as the hyperbolic solid angle of $T$ as viewed from the center point, so
this question seems strongly reminiscent of the difficult problem of proving quasiconvexity for tetrahedral mesh smoothing to maximize the minimum Euclidean solid angle, discussed in the initial subsection of this section.  A positive answer would allow the quasiconvex programming technique to be applied to this brain flat mapping application.

\subsection{Optimized Color Gamuts}

\begin{figure}[t]
\centering
\includegraphics[width=1.5in]{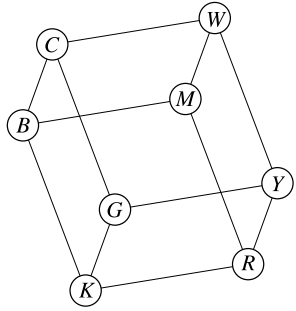}
\caption{An additive color gamut, with vertices labeled by colors:
$K={}$black, $R={}$red, $G={}$green, $B={}$blue, $C={}$cyan, $M={}$magenta, $Y={}$yellow,
$W={}$white.}
\label{fig:ColorParallelepiped}
\end{figure}

Tiled projector systems~\cite{HumHan-Vis-99,LiCheChe-CGA-00,RasBroYan-Vis-99} are a recent development in computer display technology,
in which the outputs of multiple projectors are combined into large seamless displays for
collaborative workspaces.  There are many difficult research issues involved in achieving this seamlessness: how to move the data quickly enough to all the screens, how to maintain physical alignment of the projectors, how to handle the radial reduction in brightness (vignetting) common to many projector systems, and so on.  Here we concentrate on one small piece of this puzzle:
matching colors among the outputs of multiple projectors.
Any imaging device has a {\em gamut}, the set of colors that it can produce.  However,
two projectors, even of the same model, will have somewhat different gamuts due to factors such as color filter batches and light bulb ages.  We seek a common gamut of colors that can be produced by all the projectors, and a coordinate system for that gamut so that we can display color images in a seamless fashion across multiple projectors~\cite{BerEpp-SCG-03,MajHeTow-Vis-00,Sto-CGA-01}.

Most projectors, and most computer graphics software, use an {\em additive color} system in which colors are produced by adding signals of three primary colors, typically red, green, and blue.
If we view the gamuts as sets of points in a linear three-dimensional device-independent color space, additive color systems produce gamuts that are the Minkowski sums of three line segments, one per color signal, and therefore have the geometric form of parallelepipeds
(Figure~\ref{fig:ColorParallelepiped}).
The color spaces representing human vision are three-dimensional, so these parallelepipeds have twelve degrees of freedom: three for the {\em black point} of the projector (representing the color of light it projects when it is given a zero input signal) and three each for the three primary colors (that is, the color that the projector produces when given an input signal with full strength in one primary color channel and zero in the other two color channels).
The black point and the three primary colors form four of the eight parallelepiped vertices; the other four are the secondary colors cyan, yellow, and magenta, and the white point produced when all three input color channels are saturated.

The computational task of finding a common color gamut, then, can be represented as a twelve-dim\-en\-sion\-al geometric optimization problem in which we seek the best parallelepiped to use as our gamut, according to some measure of gamut quality, while constraining our output parallelepiped to lie within the intersection of a collection of input parallelepipeds, one per projector of our system.

To represent this problem as a quasiconvex program, Bern and Eppstein~\cite{BerEpp-SCG-03} suppose that we are given eight quasiconvex functions $d_K$, $d_R$, $d_G$, $d_B$, $d_C$, $d_M$, $d_Y$, and $d_W$, where each $d_X:\R^3\mapsto\R$ measures the distance of a color from the ideal location of corner $X$ of the color cube
(here each capital letter is the initial of one of the colors at the color cube corners, except for $K$ which by convention stands for black).  This formulation allows different distance functions to be used for each color; for instance, we might want to weight $d_K$ and $d_W$ more strongly than the other six color distances.  We also form eight functions $f_X:\R^{12}\mapsto\R^3$ mapping our twelve-dimensional parametrization of color gamuts into the color values of each of the gamut corners.
If we parametrize a gamut by the black point and three primary colors, then
$f_K$, $f_R$, $f_G$, and $f_B$ are simply coordinate projections, while the other four functions are simple linear combinations of the coordinates.
For each of the eight colors $X$, define $q_X(\bar x)=d_X(f_X(\bar x))$.
The level sets of $q_X$ are simply Cartesian products of the three dimensional level sets of $d_X$ with complementary nine-dimensional subspaces of $\R^{12}$, so they are convex and each $q_X$ is quasiconvex.

It remains to formulate the requirement that our output gamut lie within the intersection of the input gamuts.  If we are given $n$ input gamuts, form a halfspace $H_{i,j}$ (with $0\le i<n$ and $0\le j<6$)
for each of the six facets of each of these parallelepipeds, and for each color $X$ form a nested convex family
$\kappa_{i,j,X}(\lambda)=\{\bar x\in\R^{12}\mid f_X(\bar x)\in H_{i,j}\}$ that ignores its argument $\lambda$ and returns a constant halfspace.
We can then represent the problem of finding a feasible gamut that minimizes the maximum distance from one of its corners to the corner's ideal location as the quasiconvex program formed by the eight quasiconvex functions $q_X$ together with the $48n$ nested convex families~$\kappa_{i,j,X}$.

\subsection{Analysis of Backtracking Recurrences}

In this section we discuss another application of quasiconvex programming, in the automated analysis of algorithms, from our paper~\cite{Epp-SODA-04-qaba}.
There has been much research on exponential-time exact algorithms for problems
that are NP-complete (so that no polynomial time solution is expected); see
\cite{Bei-SODA-99,Bys-SODA-03,DanHir-TR-00,Epp-SODA-01,Epp-WADS-01,e-arxiv,GraHirNie-SAT-00,PatPudSak-FOCS-98,Sch-FOCS-99}
for several recent papers in this area.
Although other techniques are known, many of these algorithms use a form of backtracking search
in which one repeatedly performs some case analysis to find an appropriate structure in the problem instance, and then uses that structure to split the problem into several smaller subproblems which are solved by recursive calls to the algorithm.

\begin{table*}
{\tiny $$
T(n,h) \le\max\left\{
\begin{array}{l}
T(n+3,h-2)+T(n+3,h-1)+T(n+4,h-2)+T(n+5,h-2) , \\
T(n,h+1)+T(n+1,h+2) , \\
2\,T(n+2,h)+2\,T(n+3,h) , \\
2\,T(n+2,h)+2\,T(n+3,h) , \\
T(n+3,h-2)+T(n+3,h-1)+T(n+5,h-3)+T(n+5,h-2) , \\
T(n+1,h)+T(n+3,h-1)+3\,T(n+3,h+3) , \\
T(n+3,h-2)+2\,T(n+3,h-1)+T(n+7,h-2) , \\
T(n+1,h)+2\,T(n+4,h-2) , \\
3\,T(n+1,h+2)+2\,T(n+1,h+5) , \\
2\,T(n+2,h)+T(n+3,h+1)+T(n+4,h)+T(n+4,h+1) , \\
T(n+1,h-1)+T(n+4,h-1) , \\
T(n+1,h+3)+2\,T(n+2,h)+T(n+3,h) , \\
2\,T(n+2,h-1) , \\
T(n,h+3)+T(n+1,h+2)+T(n+2,h) , \\
T(n+1,h-1)+T(n+4,h-1) , \\
2\,T(n+1,h+1)+T(n+2,h+1) , \\
9\,T(n+2,h+3) , \\
T(n+1,h)+T(n+1,h+1) , \\
9\,T(n+9,h-5)+9\,T(n+9,h-4) , \\
T(n+3,h-2)+T(n+3,h-1)+T(n+5,h-2)+2\,T(n+6,h-3) , \\
T(n+1,h-1)+T(n+4,h)+T(n+4,h+1) , \\
2\,T(n+2,h)+T(n+3,h)+T(n+4,h)+T(n+5,h) , \\
T(n+1,h)+2\,T(n+2,h+1) , \\
T(n+1,h-1) , \\
2\,T(n+2,h+1)+T(n+3,h-2)+T(n+3,h) , \\
T(n+1,h+1)+T(n+1,h+2)+T(n+2,h) , \\
2\,T(n+2,h)+2\,T(n+3,h) , \\
T(n+1,h+2)+T(n+2,h-1)+T(n+2,h+1) , \\
T(n+1,h) , \\
T(n+2,h+1)+T(n+3,h-2)+T(n+4,h-3) , \\
T(n-1,h+2) , \\
3\,T(n+4,h)+7\,T(n+4,h+1) , \\
T(n+2,h-1)+2\,T(n+3,h-1) , \\
T(n+2,h-1)+T(n+2,h)+T(n+2,h+1) , \\
T(n+3,h-2)+T(n+3,h)+2\,T(n+4,h-2) , \\
T(n+1,h)+T(n+3,h-1)+T(n+3,h+3)+T(n+5,h)+T(n+6,h-1) , \\
2\,T(n+1,h+4)+3\,T(n+3,h+1)+3\,T(n+3,h+2) , \\
3\,T(n+3,h+1)+T(n+3,h+2)+3\,T(n+3,h+3)+3\,T(n+4,h) , \\
T(n+2,h-1)+T(n+3,h-1)+T(n+4,h-2) , \\
T(n,h+1) , \\
T(n+1,h+2)+T(n+3,h-2)+T(n+3,h-1) , \\
2\,T(n+3,h-1)+T(n+3,h+2)+T(n+5,h-2)+T(n+5,h-1)+T(n+5,h)+2\,T(n+7,h-3) , \\
T(n+2,h+2)+2\,T(n+3,h)+3\,T(n+3,h+1)+T(n+4,h) , \\
T(n+3,h-2)+T(n+3,h-1)+T(n+5,h-3)+T(n+6,h-3)+T(n+7,h-4) , \\
T(n+1,h-1) , \\
T(n+1,h)+2\,T(n+3,h) , \\
4\,T(n+3,h+1)+5\,T(n+3,h+2) , \\
4\,T(n+2,h+3)+3\,T(n+4,h)+3\,T(n+4,h+1) , \\
T(n+3,h-2)+2\,T(n+3,h-1)+T(n+6,h-3) , \\
4\,T(n+2,h+3)+6\,T(n+3,h+2) , \\
T(n,h+1)+T(n+4,h-3) , \\
T(n+1,h-1)+2\,T(n+3,h+2) , \\
2\,T(n+2,h+1)+3\,T(n+2,h+3)+2\,T(n+2,h+4) , \\
2\,T(n+2,h)+2\,T(n+2,h+3) , \\
2\,T(n+2,h)+T(n+2,h+3)+T(n+3,h+2)+T(n+4,h)+T(n+4,h+1) , \\
2\,T(n,h+2) , \\
T(n+2,h)+T(n+3,h-2)+T(n+3,h-1) , \\
T(n+3,h-2)+2\,T(n+4,h-2)+T(n+5,h-3) , \\
T(n+1,h)+T(n+5,h-4)+T(n+5,h-3) , \\
T(n+1,h+2)+T(n+2,h-1)+T(n+3,h-1) , \\
T(n+2,h-1)+T(n+2,h)+T(n+4,h-1) , \\
10\,T(n+3,h+2) , \\
6\,T(n+2,h+2) , \\
T(n+2,h)+T(n+3,h) , \\
2\,T(n+3,h-1)+T(n+3,h+2)+T(n+5,h-2)+T(n+5,h-1)+T(n+5,h)+T(n+6,h-2)+T(n+7,h-2) , \\
6\,T(n+3,h+1) , \\
3\,T(n,h+3) , \\
T(n+2,h-1)+T(n+2,h)+T(n+4,h-2) , \\
2\,T(n+5,h-3)+5\,T(n+5,h-2) , \\
2\,T(n+2,h)+T(n+2,h+1)+T(n+4,h-1) , \\
8\,T(n+1,h+4) , \\
T(n+3,h-2)+T(n+3,h-1)+T(n+5,h-3)+T(n+5,h-2)+T(n+7,h-3) , \\
T(n+1,h-1)+T(n+2,h+2) , \\
5\,T(n+2,h+2)+2\,T(n+2,h+3)
\end{array}
\right.

$$}
\medskip
\caption{A recurrence arising from unpublished work with J. Byskov on graph coloring algorithms,
taken from~\cite{Epp-SODA-04-qaba}.}
\label{tbl:bigrec}
\end{table*}

For example, as part of a graph coloring algorithm~\cite{Epp-WADS-01} we used the following subroutine for listing all maximal independent sets of a graph $G$ that have at most $k$ vertices in the maximum independent set (we refer to such a set as a {\em $k$-MIS}).
The subroutine consists of several different cases, and applies the first of the cases
which is found to be present in the input graph $G$:
\begin{itemize}
\item If $G$ contains a vertex $v$ of degree zero,
recursively list each $(k-1)$-MIS in $G\setminus\{v\}$
and append $v$ to each listed set.
\item If $G$ contains a vertex $v$ of degree one, 
with neighbor $u$, recursively list each $(k-1)$-MIS in $G\setminus N(u)$ and append $u$ to each listed set.  Then, recursively list each $(k-1)$-MIS in $G\setminus \{u,v\}$ and append $v$ to each listed set.
\item If $G$ contains a path $v_1$-$v_2$-$v_3$ of degree-two vertices,
then, first, recursively list each $(k-1)$-MIS in $G\setminus N(v_1)$ and append $v_1$ to each listed set.  Second, list each $(k-1)$-MIS in $G\setminus N(v_2)$ and append $v_2$ to each listed set.  Finally, list each $(k-1)$-MIS in $G\setminus(\{v_1\}\cup N(v_3))$ and append $v_3$ to each listed set.  Note that, in the last recursive call, $v_1$ may belong to $N(v_3)$ in which case
the number of vertices is only reduced by three.
\item If $G$ contains a vertex $v$ of degree three or more,
recursively list each $k$-MIS in $G\setminus\{v\}$.
Then, recursively list each $(k-1)$-MIS in
$G\setminus N(v)$ and append $v$ to each listed set.
\end{itemize}
Clearly, at least one case is present in any nonempty graph,
and it is not hard to verify that any $k$-MIS will be generated by one of the recursive calls made from each case.  Certain of the sets generated by this algorithm as described above may not be maximal, but if these non-maximal outputs cause difficulties they can be removed by an additional postprocessing step.
We can bound the worst-case number of output sets produced by this algorithm
as the solution to the following recurrence in the variables $n$ and $k$:
$$
T(n,k)=\max\left\{
\begin{array}{l}
T(n-1,k-1)\\
2T(n-2,k-1)\\
3T(n-3,k-1)\\
T(n-1,k)+T(n-4,k-1)
\end{array}
\right.
$$
As base cases, $T(0,0)=1$, $T(n,-1)=0$, and $T(n,k)=0$ for $k>n$.
Each term in the overall maximization of the recurrence comes from a case in the case analysis; the recurrence uses the maximum of these terms because, in a worst-case analysis, the algorithm has no control over which case will arise.  Each summand in each term comes from a recursive subproblem called for that case.  It turns out that, for the range of parameters of interest $n/4\le k\le n/3$, the recurrence above is dominated by its last two terms, and has the solution $T(n,k)=(4/3)^n(3^4/4^3)^k$.  We can also find graphs having this many $k$-MISs, so the analysis given by the recurrence is tight.
Similar but somewhat more complicated multivariate recurrences have arisen in our algorithm for
3-coloring~\cite{Epp-SODA-01} with variables counting 3- and 4-value variables in a constraint satisfaction instance, and in our algorithm for the traveling salesman problem in cubic graphs~\cite{e-arxiv} with variables counting vertices, unforced edges, forced edges, and 4-cycles of unforced edges.
Another such recurrence, of greater complexity but with the same general form, is depicted in Table~\ref{tbl:bigrec}.

We would like to perform this type of analysis algorithmically: if we are given as input a recurrence such as the ones discussed above, can we efficiently determine its asymptotic solution, and determine which of the cases in the analysis are the critical ones for the performance of the backtracking algorithm that generated the recurrence?  We showed~\cite{Epp-SODA-04-qaba} that these questions can be answered automatically by a quasiconvex programming algorithm, as follows.

Let $\bar x$ denote a vector of arguments to the input recurrence,
and for each term in the input recurrence define a univariate linear recurrence,
by replacing $\bar x$ with a weighted linear combination $\xi=\bar w\cdot\bar x$ throughout.
For instance, in the $k$-bounded maximal independent set recurrences, the four terms in the recurrence
lead to four linear recurrences
$$
\begin{array}{l}
t_1(\xi) = t_1(\xi - \bar w\cdot(1,1)) \\
t_2(\xi) = 2t_2(\xi - \bar w\cdot (2,1)) \\
t_3(\xi) = 3t_3(\xi - \bar w\cdot (3,1)) \\
t_4(\xi) = t_4(\xi - \bar w\cdot (1,0)) + t_4(\xi - \bar w\cdot (4,1))
\end{array}.
$$
We can solve each of these linear recurrences to find constants $c_i$
such that $t_i(\xi)=O(c_i^\xi)$; it follows that, for any weight vector $\bar w$, $T(\bar x)=O(\max c_i^{\bar w\cdot\bar x})$.

This technique only yields a valid bound when each linear recurrence is solvable; that is,
when each term on the right hand side of each linear recurrence has a strictly smaller argument
than the term on the left hand side.
In addition, different choices of $\bar w$ in this upper bound technique will give us different bounds.

To get the tightest possible upper bound from this technique, for $\bar x=n\bar t$ where $\bar t$
is a fixed {\em target vector}, constrain $\bar w\cdot\bar t=1$ 
(this is a normalizing condition since multiplying $\bar w$ by a scalar does not
affect the overall upper bound), and
express $c_i$ as a function $c_i=q_i(\bar w)$ of the weight vector~$\bar w$;
set $c_i=+\infty$ whenever the corresponding linear inequality has a right hand side term with argument greater than or equal to that on the left hand side.
We show in~\cite{Epp-SODA-04-qaba} that these functions $q_i$ are quasiconvex, as
their level sets can be expressed by the formula
$$q_i^{\le\lambda}=\Bigl\{\bar w \Bigm{|} \sum_j \lambda^{-\bar w\cdot\delta_{i,j}}\le 1\Bigr\},$$
where the right hand side describes a level set of a sum of convex functions of~$\bar w$.
Therefore, we can find the vector $\bar w$ minimizing
$\max_i q_i(w)$ as a quasiconvex program.  The value $\lambda$ of this quasiconvex program
gives us an upper bound $T(n\bar t) = O(\lambda^n)$ on our input recurrence.

In the same paper, we also show a lower bound
$T(n\bar t) = \Omega(\lambda^n n^{-c})$,
so the upper bound is tight to within a factor that is polylogarithmic compared to the overall solution.
The lower bound technique involves relating the recurrence solution to the probability that a random walk in a certain
infinite directed graph reaches the origin, where the sets of outgoing edges from each vertex in the graph are also determined randomly with probabilities determined from the gradients surrounding the optimal solution of the quasiconvex program for the upper bound.

\subsection{Robust Statistics}

If one has a set of $n$ observations $x_i\in\R$, and wishes to summarize them by a single number, the average or mean is a common choice.  However, it is sensitive to {\em outliers}: replacing a single observation by a value far from the mean can change the mean to an arbitrarily chosen value.  In contrast, if one uses the median in place of the mean, at least $n/2$ observations need to be corrupted before the median can be changed to an arbitrary value; if fewer than $n/2$ observations are corrupted, the median will remain within the interval spanned by the uncorrupted values.  In this sense, the median is {\em robust} while the mean is not.  More generally, we define a statistic to be robust if its {\em breakdown point} (the number of observations that must be corrupted to cause it to take an arbitrary value) is at least $cn$ for some constant $c>0$.

\begin{figure}[t]
\centering
\includegraphics[width=2.5in]{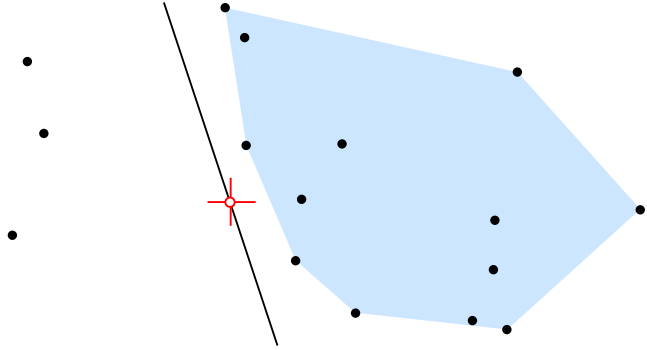}
\caption{The Tukey depth of the point marked with the $+$ sign is three: there is a halfplane containing it and only three sample points (shown as solid disks); or, equivalently, three points can be removed from the sample set to place the test point outside the convex hull of the remaining points (shaded).}
\label{fig:TukeyDepth}
\end{figure}

If one has observations $\bar x_i\in\R^d$, it is again natural to attempt to summarize them by a single point $\bar x\in\R^d$.  In an attempt to generalize the success of the median in the one-dimensional problem, statisticians have devised many notions of the {\em depth} of a point, from which we can define a generalized median as being the point of greatest depth~\cite{GilSteWig-DM-92,Hod-AMS-55,Liu-AS-90,LiuParSin-AS-99,Mah-PNASI-36,Oja-SPL-83,Tuk-ICM-75,ZuoSer-AS-00}.
Of these definitions, the most important and most commonly used is the {\em Tukey depth}~\cite{Hod-AMS-55,Tuk-ICM-75}, also
known as {\em halfspace depth} or {\em location depth}.
According to this definition, the depth of a point $\bar x$ (which need not be one of our sample points)
is the minimum number of sample points contained in any halfspace that contains~$\bar x$
(Figure~\ref{fig:TukeyDepth}).
The {\em Tukey median} is any point of maximum depth.  It follows by applying Helly's theorem to the system of halfspaces containing more than $dn/(d+1)$ observations that, for observations in $\R^d$, the Tukey median must have depth at least $n/(d+1)$. This depth is also its breakdown point, so the Tukey median is robust, and it has other useful statistical properties as well, such as invariance under affine transformations and the ability to form a center-outward ordering of the observations based on their depths.

There has been much research on the computation of Tukey medians, and of other points with high Tukey depth~\cite{Cha-SODA-04,ClaEppMil-IJCGA-96,Col-JACM-87,ColShaYap-SJC-87,JadMuk-DCG-94,LanSte-JCDCG-00,LanSte-ICORS-01,LanSte-STACS-03,Mat-DCG-92,NaoSha-CCCG-90,RouRut-SS-98,StrRou-CSDA-00}.  Improving on many previously published algorithms,
Chan~\cite{Cha-SODA-04} found the best bound known for Tukey median construction,
$O(n\log n + n^{d-1})$ randomized expected time, using his implicit quasiconvex programming technique.

Let $B$ be a bounding box of the sample points.
Each $d$-tuple $t$ of sample points that are in general position in $\R^d$ defines a hyperplane that bounds two closed halfspaces, $H_t^+$ and $H_t^-$.  If we associate with each such halfspace a number $k_t^+$ or $k_t^-$ that counts the number of sample points in the corresponding halfspace, then the pairs $(B\cap H_t^\pm,-k_t^\pm)$ can be used to form a generalized longest intersecting prefix problem, as defined in Section~\ref{sec:lip}; borrowing the terminology of LP-type problems, call any such pair a {\em constraint}. The solution to the quasiconvex program defined by this set of constraints is a pair $(k,\bar x)$ where $k$ is minimal and every halfspace with more than $k$ samples contains $\bar x$.
If a halfspace $H$ contains fewer than $n-k$ samples, therefore, it does not contain $\bar x$,
so the depth of $\bar x$ is at least $n-k$.  Any point of greater depth would lead to a better solution to the problem, so $\bar x$ must be a Tukey median of the samples, and we can express the problem of finding a Tukey median as a quasiconvex program.  This program, however, has $O(n^d)$ constraints, larger than Chan's claimed time bound.
To find Tukey medians more quickly, Chan applies his implicit quasiconvex programming technique: we need to be able to solve constant sized subproblems in constant time, solve decision problems efficiently, and partition large problems into smaller subproblems.

It is tempting to perform the partition step as described after Theorem~\ref{thm:iqcp},
by dividing the set of samples arbitrarily into $d+1$ equal-sized subsets and using the complements of these subsets.  However, this idea does not seem to work well for the Tukey median problem: the difficulty is that
the numbers $k_t^\pm$ do not depend only on the subset, but on the whole original set of sample points.

Instead, Chan modifies the generalized longest intersecting prefix problem (in a way that doesn't change its optimal value) by including a constraint for every possible halfspace, not just those halfspaces bounded by $d$-tuples of samples.
There are infinitely many such constraints but that will not be problematic as long as we can satisfy the requirements of the implicit quasiconvex programming technique.  To perform the partition step for this technique,
we use a standard tool for divide and conquer in geometric algorithms, known as $\epsilon$-cuttings.  We form the projective dual of the sample points, which is an arrangement of hyperplanes in $\R^d$; each possible constraint boundary is dual to a point in $\R^d$ somewhere in this arrangement,
and the number $k_t^\pm$ for the constraint equals the number of arrangement hyperplanes above or below this dual point.   We then partition the arrangement into a constant number of simplices, such that each simplex is crossed by at most $\epsilon n$ hyperplanes.  For each simplex we form a subproblem,
consisting of the sample points corresponding to hyperplanes that cross the simplex, together with a constant amount of extra information: the simplex itself and the numbers of hyperplanes that pass above and below it.  Each such subproblem corresponds to a set of constraints dual to points in the simplex.
When recursively dividing a subproblem already of this form into even smaller sub-subproblems, we intersect the sub-subproblem simplices with the subproblem simplex and partition the resulting polytopes into smaller simplices; this increases the number of sub-subproblems by a constant factor.  In this way we fulfill the condition of Theorem~\ref{thm:iqcp} that we can divide a large problem into a constant number of subproblems, each described by an input of size a constant fraction of the original.

Subproblems of constant size may be solved by constructing and searching the arrangement dual to the samples within the simplex defining the subproblem.
It remains to describe how to perform the decision algorithm needed for Theorem~\ref{thm:iqcp}.
Decision algorithms for testing the Tukey depth of a point were already known~\cite{RouRut-JRSSC-96,RouStr-SC-98}, but here we need to solve a slightly more general problem due to the extra information associated with each subproblem.
Given $k$, $\bar x$, and a subproblem of our overall problem, we must determine whether
there exists a {\em violated constraint}; that is, a halfspace that is dual to a point in the simplex defined by the subproblem,
and that contains more than $k$ sample points but does not contain $\bar x$.
Let $H$ be the hyperplane dual to $\bar x$, and $\Delta$ be the simplex defining the subproblem.
If there exists a violated constraint dual to a point $h\in\Delta$,
we can assume without loss of generality that either $h\in H$ or $h$ is on the boundary of $\Delta$;
for, if not, we could find another halfspace containing as many or more samples
by moving $h$ along a vertical line segment until it reaches either $H$ or the boundary.
Within $H$ and each boundary plane of the simplex, we can construct the $(d-1)$-dimensional arrangement formed by intersecting this plane with the planes dual to the sample points, in time $O(n\log n + n^{d-1})$. Within each face of these arrangements, all points are dual to halfspaces that contain the same number of samples, and as we move from face to face, the number of sample points contained in the halfspaces changes by $\pm 1$, so we can compute these numbers in constant time per face as we construct these arrangements.  By searching all faces of these arrangements we can find a violated constraint, if one exists.

To summarize, by applying the implicit quasiconvex programming technique of Theorem~\ref{thm:iqcp} to a generalized longest intersecting prefix problem, using $\epsilon$-cuttings to partition problems into subproblems and $(d-1)$-dimensional arrangements to solve the decision algorithm as described above, Chan~\cite{Cha-SODA-04} shows how to find the Tukey median of any point set in randomized expected time $O(n\log n + n^{d-1})$.

\section{Conclusions}

We have introduced quasiconvex programming as a formalization for geometric optimization intermediate in expressivity between linear and convex programming on the one hand, and LP-type problems on the other. 
Quasiconvex programs are capable of expressing a wide variety of geometric optimization problems and applications, but are still sufficiently concrete that they can be solved both by rapidly converging numeric local improvement techniques and (given the assumption of constant-time primitives for solving constant-sized subproblems) by strongly-polynomial combinatorial optimization algorithms.   The power of this approach is demonstrated by the many and varied applications in which quasiconvex programming arises.

\section*{Acknowledgements}

This research was supported in part by NSF grant CCR-9912338.
 
\raggedright
\bibliographystyle{abuser}
\bibliography{qcp}
 \end{document}